\documentstyle[12pt]{article}
\input mssymb

\newtheorem{theorem}{Theorem}
\newtheorem{prop}{Proposition}
\newtheorem{lemma}{Lemma}
\newtheorem{conj}{Conjecture}
\newcommand{\PB}{\stackrel{\textstyle\otimes}{,}}

\def\ma#1#2#3#4{\left[{}^{#1}_{#3}{}^{#2}_{#4}\right]}

\begin{document}

\title{On the Two Gap Locus for the Elliptic Calogero--Moser Model}
\author{ V Z Enolskii${}\sp{1,2}$, J C Eilbeck${}\sp 1$\\ ${}\sp
1$Department of Mathematics, Heriot-Watt University\\ Riccarton,
Edinburgh EH14 4AS, Scotland\\ ${}\sp 2$ Department of Theoretical
Physics\\ Institute of Metal Physics \\ Vernadsky str. 36, Kiev-680,
252142, Ukraine}

\maketitle

\begin{abstract} We give an analytical description of the locus of the
two-gap elliptic potentials associated with the corresponding flow of
the Calogero--Moser system. We start with the description of
Treibich--Verdier two--gap elliptic potentials. The explicit formulae
for the covers, wave functions and Lam\'e polynomials are derived,
together with a new Lax representation for the particle dynamics on the
locus.  Then we consider more general potentials within the Weierstrass
reduction theory of theta functions to lower genera. The reduction
conditions in the moduli space of the genus two algebraic curves are
given. This is some subvariety of the Humbert surface, which can be
singled out by the condition of the vanishing of some theta constants.
\end{abstract}
\newpage

\section{Introduction} The Calogero-Moser model, whose complete
integrability was shown a number of years ago (c.f.\cite{op81}),
continues to attract more and more attention.  This model has a rich
algebraic-geometrical structure: its flows are connected with the pole
dynamics of elliptic solutions of completely integrable partial
differential equations \cite{amm77}, the Lax representation for the
model depends through elliptic functions on the spectral parameter
\cite{kr80} and only the integration in terms of zeros of theta
functions for the model is known \cite{kr80}. The system permits a
relativistic generalisation, which also is completely integrable
\cite{rs86}.

The classical Poisson  $r$-matrix structure for the elliptic
Calogero-Moser model was described very recently \cite{bs93,skl93}. The
$r$-matrix found appears to be linear of dynamical type, i.e. dependent
om the dynamical variables.  The classical Poisson structure for the
relativistic generalisation of the Calogero-Moser model is described
only in the soliton case with a quadratic $r$-matrix of dynamical type
\cite{bb93}. The separated variables for these systems remain an
unsolved problem besides the case of a small number of particles
particles (see, e.g.\ \cite{br94}).

The quantum Calogero-Moser problem has also a rich algebraic structure
\cite{fv93}. It is remarkable that the solutions of the quantum problem
are isomorphic to the solutions of the Knizhnik-Zamolodchikov equations
which are now understood to play an important role in the theory of
quantum integrable models \cite{smf92a}.

Because the Calogero-Moser model describes the pole dynamics for the
elliptic solutions of the Kadomtsev-Petviashvili type equations
\cite{amm77}, its elliptic case becomes the classically known Lam\'e
potentials of the Schr\"odinger equation. Although this paper is
devoted to the investigations of elliptic potentials of the
one-dimensional Schr\"odinger equation, we emphasise the importance of
such potentials for different problems: the finite gap multidimensional
spectral problem \cite{vsc93}, Wess-Zumino-Witten model on the torus
\cite{ek93a} and others.

All the results given below can be generalised to higher genera, but we
shall restrict ourselves to the investigation of the first nontrivial
case of genus two to give the more complete formulae.  Through all the
paper we use computer algebra systems (Mathematica \cite{wo91} and
Maple \cite{maple92}) to derive and simplify the formulae.

The paper is organized as follows. In Section 2 we discuss the linear
$r$-matrix algebra for the Calogero-Moser system and define its
restriction to the locus associated with the KdV dynamics. In Section 3
we describe the two-gap Lam\'e and Treibich-Verdier potentials
\cite{tv90,tv91} for which we find explicitly the covers over the tori,
derive the wave functions of the associated Schr\"odinger equations and
Lam\'e polynomials.  We also give a new Lax representations for the
dynamics of particles on the locus in terms of $2\times2$ matrices. We
show in Section 4 that Treibich-Verdier potentials are some special
cases of elliptic potentials.  Using the classical reduction theory of
Riemann theta functions to lower genera (see e.g.\
\cite{bbeim94,be93,kr03}), we give necessary and sufficient conditions
under which the two-gap potential is  elliptic. We formulate these
conditions in terms of vanishing of some theta constants which in  turn
are some subvarieties of Humbert surfaces (see
e.g.\ \cite{kr03,ge82}).  We derive one of the two-gap Treibich-Verdier
potentials from this theta functional approach and give a new example
of an elliptic potential.  The paper is supported by two  appendices
which contains the description of spectral characteristics of the
Treibich-Verdier potentials and all the necessary formulae to complete
the theta functional computations mentioned in the paper.

\section{The Calogero-Moser System on the Locus}

The elliptic Calogero-Moser model is the system of $N$ one-dimensional
particles interacting via a two-particle potential described by
the Hamiltonian

\begin{equation}
  H=\sum_i y^2_i + \sum_{i,j} \wp(x_{i}-x_{j}), \label{H}
\end{equation}
with $\wp$ being the Weierstrass elliptic function \cite{Ba55} with the
periods $2\omega_1$, $2\omega_2$ and $y_i$, $x_i$ ($u=1,\ldots,N$)
being canonical variables, $\{y_{i}, y_{j}\}=\{x_i, x_j\}=0$, $\{y_i,
x_j\}=\delta_{ij}$.

Let $\{X_\mu\}=\{H_i,E_\alpha\}$, be basis matrices,
$H_i=(\delta_{ij}\delta_{ik}),i=1,\ldots,N$,
$E_{\alpha}=E_{nm}=(\delta_{nj}\delta_{nk}), n\neq m, m,n=1,\ldots,N$.

The Lax operator of the system $L$ was found by Krichever and has the
form \cite{kr80} \begin{equation}
 L(u)=\sum_{j}  y_jH_j+i\sum_{\alpha} \Phi_{\alpha}E_{\alpha},
\label{eq:defL}\end{equation}
where
\begin{equation}\Phi_{\alpha}=\Phi({\bf x}\cdot \alpha;u),\quad
\Phi(x;u)={\sigma(x-u)\over \sigma(x)\sigma(u)} {\rm
e}^{\zeta(u)x},\label{phi}\end{equation}
 where $\sigma$ and $\zeta$ are Weierstrass functions. The Hamiltonian
flows of the system are generated by ${\rm Tr}\, L^n$, in particular,
${\rm Tr} \,L^2$ gives the Hamiltonian (\ref{H}).

The Poisson structure of the system, as recently shown by
Sklyanin \cite{skl93} and Braden and Suzuki \cite{bs93}, is described
by a linear dynamical $r$-matrix algebra,

\begin{equation} \{L_{1}(u), L_{2}(v)\}=[r_{12}(u,v),
L_{1}(u)]-[r_{21}(u,v), L_{2}(v)], \label{linal} \end{equation} where
$L_1= L\otimes I, \quad L_2=I\otimes L$,
$r_{12}(u,v)=\sum_{\mu,\nu}r^{\mu\nu}(u,v)X_{\mu}\otimes X_{\nu}$ is an
$N^2\times N^2$ matrix depending on the dynamical variables, and
$r_{21}(u,v)= P r_{12}(v,u) P,$ $P$ is the permutation:  $Px\otimes
y=y\otimes x$.  The nonzero elements of the $r$-matrix are \cite{bs93}
\begin{eqnarray} &&r^{-\alpha\alpha}(u,v)=\Phi_{\alpha}(v-u){\rm
e}^{\psi(u,v)},\quad r^{i\alpha}(u,v)={1\over2}\Phi_{\alpha}(v),
\nonumber\\&&r^{ij}(u,v)=\psi(u,v)\delta_{ij},\label{rmatrix}\\
&&\psi(u,v)=\zeta(v-u)+\zeta(u)-\zeta(v),\nonumber \end{eqnarray}
satisfying the dynamical Yang-Baxter equation, \begin{eqnarray}
&&[d_{12}(x,y),d_{13}(x,z)]+[d_{12}(x,y),d_{23}(y,z)]
+[d_{32}(z,y),d_{13}(x,z)]\nonumber\\&&+ \{L_2(y)\PB
d_{13}(x,z)\}-\{L_3(z)\PB d_{12}(x,y)\} \nonumber\\
&&+[S_{13}(x,z),L_2(y)]-[S_{12}(x,y),L_3(z)]=0,\label{dybe}
\end{eqnarray} where the two other equations are obtained by cyclic
permutations and in this context $d_{12}(x,y)=r_{12}(x,y)\otimes I$,
$d_{23}(y,z)=I\otimes r_{23}(y,z)$,
$d_{13}(x,y)=\sum_{\mu,\nu}r^{\mu\nu}(x,y)X_{\mu}\otimes I\otimes
X_{\nu}$, and the $S$-matrix has the form \begin{eqnarray}
&&S_{13}(x,z)=-\Phi_{\alpha}(x-z){\rm exp} \,\psi(x,z)
E_{-\alpha}\otimes H_i\otimes E_{\alpha},\nonumber\\
&&S_{12}(x,y)=-\Phi_{\alpha}(x-y){\rm exp}\, \psi(x,y)
E_{-\alpha}\otimes E_{\alpha}\otimes H_i.\label{tail} \end{eqnarray}

We point out that the $S$-matrix (\ref{tail}) differs from that given
by Sklyanin \cite{skl93}, where a different representation for the
operator $L$ was considered. Although the $S$-term  appeared already in
\cite{eekl93aa}, its significance became more evident after
\cite{skl93}.

The equation \begin{equation} {\rm det}( L-\lambda I )=0\label{krcurve}
\end{equation} defines the {\it Krichever curve} i.e. the algebraic
curve $C_N=(k,u)$ which is an $N$-sheeted cover of a torus in $\pi:
C_N\rightarrow C_1$ \begin{equation} \lambda^N+\sum_{i=0}^{N-1}r_i(u)
\lambda^{N-i-1}=0, \end{equation} where $r_i(u)$ are elliptic
functions. In particular, for the first two of them we have
$r_1(u)=-{N\choose 2}\wp(u)+\sum \wp_{ij}$, $r_2(u)={N\choose
3}\wp^{\prime}(u)$.

We consider the restriction of the third flow, ${\rm Tr}\,L^3$ of the
Calogero-Moser system to the variety of stable points of the second
flow, ${\rm grad} \, H$ - the {\it locus} ${\cal L}_N$ \begin{equation}
{\cal L }_N = \left\{({\bf x}, {\bf y})\big|y_i=0,\quad \sum_{i\ne j}
\wp'(x_i - x_j) = 0, \,x_i \ne x_j, \, i, j=1,\ldots,N \right\}.
\label{locus} \end{equation}

It is shown in \cite{amm77} that if the particles $x_i$ move over the
locus according to the equation \begin{equation}{d x_i\over dt
}=-12\sum_{j=1, j\neq i}^N \wp(x_i-x_j),\quad
i=1,\ldots,N,\label{flow}\end{equation} then \begin{equation} u(x) = 2
\sum_{j=1}^{N} \wp(x-x_j(t)) + C, \label{elsol} \end{equation} is an
elliptic solution of the KdV equation $u_t=6uu_x-u_{xxx}$ where $C$ is
a constant.

The geometry of the locus ${\cal L}_N $ was studied by Airault, McKean
and Moser \cite{amm77} and others.  They showed that  the locus is
nonempty for positive triangle  integers $N$, i.e.\ for  numbers of the
form  $N = g(g+1)/2$, where $g$ is the number of gaps in the spectrum
(or the genus of the corresponding algebraic curve).  The corresponding
elliptic potential is the $g$-gap Lam\'e potential.  Recently Treibich
and Verdier \cite{tv91} found a new set of elliptic potentials of the
form (\ref{elsol}) corresponding to nontriangle numbers of points on
the locus ${\cal L}_N$. In particular for the points of the locus
$x_i$ being the half-periods they found a family of elliptic potentials
of the form \cite{tv90} \begin{equation} u(x)=\sum_{i=0}^3{g_i(g_i+1)}
\wp(x-\omega_i), \quad g_i\in  {\Bbb N},\label{tvp} \end{equation}
which are associated with the cover of degree
$N={1\over2}\sum_{i=0}^3g_i(g_i+1)$ over a torus. We shall refer to
these potentials as {\it Treibich-Verdier potentials}.

The curve (\ref{krcurve}) becomes  hyperelliptic when restricted to the
${\cal L}_N$ \cite{bbeim94}. Therefore one expects to be able to write
down a $2\times2$ Lax representation for the particle dynamics on a
locus.  This is done below for the two-gap Lam\'e and Treibich-Verdier
potentials.  Nevertheless the $r$-matrix formulation of the
Calegero-Moser flows restricted to the locus remains an unsolved
problem.

\section{Two-gap Treibich-Verdier Potentials} \subsection{The spectral
characteristics of elliptic solitons } We shall start on the potential
of the form (\ref{tvp}).  There exist exactly 6 two-gap
Treibich-Verdier potentials $u_N(x)$ associated with $N$-sheeted
covering of the torus.  \begin{table}[htpb] \caption{6 Two-gap
Treibich-Verdier potentials} \begin{tabular}{||l|l||}\hline
$N$&$u_N(x)$\\ \hline 3&$6\wp(x)$ or
$2\wp(x+\omega_1)+2\wp(x+\omega_2)+2\wp(x+\omega_3)$\\ \hline
4&$6\wp(x)+2\wp(x+\omega_i), \; i=1,2,3$\\ \hline
5&$6\wp(x)+2\wp(x+\omega_i)+2\wp(x+\omega_j)\;i,j=1,2,3$\\ \hline
6&$6\wp(x)+6\wp(x+\omega_i)\; i= 1,2,3$\\ \hline
8&$6\wp(x)+6\wp(x+\omega_i)+2\wp(x+\omega_j)+2\wp(x+\omega_k)\;
i,j,k=1,2,3$\\ \hline 12&$6\wp(x)+6\wp(x+\omega_1)+6\wp(x+\omega_2)
+6\wp(x+\omega_3)$\\  \hline \end{tabular} \end{table}

We note that the three last potentials are simply the two order
transformation (Gauss transformation) of the first two potentials
\begin{equation} \wp\left(z|\omega, {1\over
2}\omega'\right)=\wp(z)+\wp(z+\omega')\label{gauss} \end{equation}
Therefore we shall refer to the first three potentials as {\it
primitive}.

To describe the two gap Lam\'e potential $6\wp(x)$ and primitive
Treibich-Verdier potentials we have to \begin{itemize} \item exhibit
the associated algebraic curve of genus two \begin{equation}
C_2=(w,z),\,w^2=\prod_{i=1}^5(z-z_i) \label{curve} \end{equation}
\item  give its covers $\pi: C_2\rightarrow C_1$ and $\widetilde\pi:
C_2\rightarrow \widetilde C_1$ over the tori $C_1=(\wp',\wp),
\;(\wp')^2=4\wp^3-g_2\wp-g_3$ and $ \widetilde
C_1=(\widetilde\wp',\widetilde\wp),\;
(\widetilde\wp')^2=4\widetilde\wp^3-\widetilde
g_2\widetilde\wp-\widetilde g_3,$ where the moduli $\widetilde
g_2,\,\widetilde g_3$ are expressed in some way through the moduli
$g_2,g_3$.  \item  describe the two-gap locus  ${\cal L}_N$ \item
write the solution $\Psi$ of the Schr\"odinger equation.  \end{itemize}
We can do all this by  classical means (which modern computer algebra
makes more effective) following the work of Hermite \cite{he12} and
Halphen \cite{ha84}.

Let us consider the Lam\'e equation \begin{equation} \left[{\partial^2
\over \partial x^2 } - \sum_{i=1}^n a_i(a_i+1)\wp(x-x_i)\right]
\Psi(x;u)= z\Psi(x;u), \label{lame} \end{equation} where
$\sum_{i=1}^n{1\over2}a_i(a_1+1)=N$ is the degree of the cover.

We shall use the following generalisation of the Hermite \cite{he12}
and Halphen \cite{ha84} ansatz for the function $\Psi$,
\begin{equation} \Psi(x;u) = e^{kx}\sum_{i=1}^n \sum_{j=0}^{a_j-1}
A_j(z,k,u) {\partial^{j} \over \partial
 x^{j}}\Phi{(x-x_i,u)}, \label{ans} \end{equation} where the function
$\Phi(x;u)$ is the solution of (\ref{lame}) for $n = 1,\, a_1=1$ is
given by (\ref{phi}) and $A_j(z,k,u)$ are some functions of the
spectral parameters  $z, k, u$. Although the ansatz is valid for any
point of the locus ${\cal L}_N$ we shall consider below only  special
points of the form $x_i=\omega_i$ or $0$ found in \cite{tv90} and
listed above in  Table 1.  We shall refer to the {\it Lam\'e\/}
polynominals $\Lambda_k(x)$ as the values of $\Psi(x;u)$ at values of
$u$ corresponding to the edges of the gaps $u=u_k, k=1,\dots,5$.

After substituting the expansions of $\Phi(x,u)$ near the pole at
$x=0$, \begin{equation} \Phi(x,u) = {1\over x} - {{\wp(u)}\over 2} x +
{{\wp'(u)}\over 6} x^2 +
	  {{g_2-5\wp^2(u)}\over{40}} x^3+ \ldots \label{1.15}
\end{equation} and near $x=\omega_i$ \begin{eqnarray}
\Phi(x+\omega_i)=  \Phi(\omega_i)\left(1+{\wp'(u)\over 2(e_i-
\wp(u))}x+{1\over2}(2e_i+\wp(u))x^2+\ldots\right) \label{1}
\end{eqnarray} to (\ref{lame}) and equating the principal parts of the
poles we come to an overdetermined linear system   for the $A_i$. The
compatibility conditions give exactly two conditions \begin{equation}
{\cal P}_1(k,z,\wp(u))=0,\quad {\cal P}_2(k,z,\wp(u))=0\label{compat}
\end{equation}
 with polynomials ${\cal P}_i$ of their arguments. By eliminating the
 variables $z$ or $\wp$ from the conditions (\ref{compat})  we obtain
two equivalent realizations of the curve (\ref{curve}); eliminating the
variable $z$ we obtain the first cover.

To find the second cover we use the fact that there exists the
reduction formula \begin{equation} {dz\over w}=-{d\widetilde \wp\over
\widetilde \wp'}\label{sec} \end{equation} with $(\widetilde
\wp^{\prime},\widetilde \wp)$ lying on the torus $\widetilde C_1$ and
the coordinate $\widetilde \wp$ being a rational function of $z$,
\begin{equation} \wp= {Q_N(z)\over P_{N-3}(z)},\label{rat}
\end{equation} where $Q_N$ and $P_{N-3}$ are polynomials of orders $N$
and $N-3$ respectively.

The description of the spectral characteristics of the primitive
Treibich-Verdier potentials is given in the Appendix A.

\subsection{The Dynamics on the Locus} The complete description of the
dynamics on the locus under the action of the KdV flow was given in
\cite{amm77} for the case of the two-gap Lam\'e potential by some
tricky manipulations with equations (\ref{locus}) and (\ref{flow}). It
was shown that the dynamics are described by a foliation where the
basis  and the bundle are respectively the elliptic curves $C_1$ and
$\tilde C_1$ whose moduli are inter-dependent. The paper also
conjectured that the  same foliation  would occur for all two-gap
elliptic potentials.

We show below how to compute the second curve $\tilde C_1$ for
primitive Trei-\break bich-Verdier potentials. The statement of
\cite{amm77} about the foliation can be proved by means of the
Weierstrass reduction theory \cite{bbeim94,kr03} in the next section.

To describe the dynamics over the locus we write the Jacobi inversion
problem, for the curve associated with elliptic potential
\begin{eqnarray} \int_{\infty}^{\mu_1} {zdz\over w} +
\int_{\infty}^{\mu_2} {zdz\over w} =2ix+C_1,\quad \int_{\infty}^{\mu_1}
{dz\over w} + \int_{\infty}^{\mu_2} {dz\over w} =-8it +C_2\label{4.32}
\end{eqnarray} From the {\it trace formulae} \cite{zmnp80} written for
the elliptic potential in the form \begin{eqnarray} \mu_1+\mu_2 &=&
-\sum_{j=1}^N \wp (x - x_j) + {1\over 2} \sum_{j=1}^5 z_j,\nonumber \cr
\mu_1 \mu_2 &=&3\,\sum_{i<j} \wp(x - x_i) \wp(x -x_j) -  {Ng_2\over8}
+{1\over 2} \sum_{i<j} z_i z_j - {3\over 8}\left(\sum_{j=1}^5 z_j
\right)^2 \label{4.33} \end{eqnarray} we find in the vicinity of the
point $x_j$ the decompositions \begin{eqnarray} \mu_1 (x_j +
\epsilon,t) = {1\over {\epsilon}^2} + o (1), \,\, \mu_2
(x_j+\epsilon,t) = - 3 \left(\sum_{i\not=j} {\wp(x_j-x_i)}
\right)+o(\epsilon) \label{4.34} \end{eqnarray}

Therefore the equations (\ref{4.32}) in which $x=x_j$ and integrals are
hyperelliptic are expressed in terms of elliptic functions in the
following way \begin{eqnarray} {\cal Q}_1 (\mu_1 (x_j)) = \wp (ax_j +bt
+c),\quad {\cal Q}_2 (\mu_1 (x_j)) = \widetilde {\wp}(dt + e),
\label{4.35} \end{eqnarray} where ${\cal P}_{1,2}$ are rational
functions of the $N$--th degree, $\wp$ and $\widetilde \wp$ are
Weierstrass elliptic functions defined on the first and second tori
respectively; $a,\,b,\,c,\,d,\,e$ are constants that appear under
reduction. By eliminating  the variable $\mu_1$ from (\ref{4.34}), we
have an algebraic equation of the $N$--th degree with respect to $\wp$
and coefficients depending on $\widetilde \wp$.

In particular, we have the following  isospectral deformation of the
potentials $u_3$, $u_4$ and $u_5$. Let \[ X_j=-3\sum_{k\neq
j}^N\wp(x_j-x_k),\quad j=1,\ldots,N.  \] Then we have for $N= 3,4,5$
respectively \begin{eqnarray}
u_3:&&4X^3-9g_2X+9g_3+{16\over9}\tilde\wp(8it)=0,\label{ev3}\\
u_4:&&9(X-z_2)(X-z_3)(X+4e_i-e_k)^2\nonumber\\&&\qquad+
4(X+6e_i)(\tilde\wp(8{\rm i}t)-\tilde e_j)=0,\label{ev4}\\
u_5:&&9P_5(X)+4(X-3e_i-9e_j)(X-3e_i-9e_k)\tilde\wp(8{\rm
i}t)=0,\label{ev5} \end{eqnarray} where the polynomial $P_5(z)$ in
(\ref{ev5}) is given in Table 4 in the Appendix A.

We note that the rational limit of the  dynamics is the same for all
the potentials. The equations (\ref{ev3}-\ref{ev5}) give the
integration of the corresponding Calogero-Moser flows restricted to the
locus.

We also note that  equation (\ref{ev3}) can be extracted from
\cite{amm77}, p.144.

Let us construct the Lax representation for Calogero-Moser system,
being restricted to the locus. We choose the ansatz for such a
representation in the form of $2\times 2$ matrices \begin{eqnarray}
\dot L(z)&=&[M(z),L(z)],\nonumber\\ L(z)&=&\left(\matrix{ V(z)&U(z)\cr
W(z)&-V(z)}\right), \quad M(z)=\left(\matrix{ 0&1\cr
Q(z)&0}\right).\label{Lax}\end{eqnarray} It follows from the equation
(\ref{Lax}) that \begin{eqnarray}&&V(z)=-\frac12 \dot{U}(z), \quad
W(z)=-\frac12 \ddot{U}(z)+U(u)Q(z),\nonumber\\ &&\dot
W(z)=2V(z)Q(z).\label{Lax1} \end{eqnarray}

To construct the Lax representation we have to define $U(z)$ and
$Q(z)$. Let us introduce the following ansatz \begin{eqnarray}
U(z)=\prod_{i=1}^N (z-X_i),\quad Q(z)=\zeta+2\tilde\wp(8 i
t),\label{defuq} \end{eqnarray} where the polynomials $U(z)$ and the
function $\zeta$ is the expression for the second cover taken from
Table 4 and the quantity $\tilde\wp(8it)$ is expressed in terms of
$X_i$ from the equations (\ref{ev3}-\ref{ev5}) with the help of the
Viett theorem.

The spectral curve has the form \begin{equation} Y^2=w^2(z)
\left({\partial \tilde\wp\over \partial z}\right)^2, \end{equation}
where the polynomial $w^2$ is taken from  Table 2 and $\tilde\wp$ is
the rational function taken from  Table 4.

To find these Lax representations we use the Lax representation for the
dynamics associated with the curve $\tilde C_1$, with \begin{equation}
U(\zeta)=\zeta-\tilde\wp(8it), \quad Q(\zeta)= \zeta+2\tilde\wp(8it)
\end{equation} and raise this representation to the curve $C_2$ using
the formulae for the cover.

For example,  let us consider the particle dynamics associated with the
two-gap Lam\'e potential $u_3$ which is described by the equations
\begin{equation} \wp'_{12}+\wp'_{13}=0,\;\wp'_{21}+\wp'_{23}=0,
\;\wp'_{31}+\wp'_{32}=0,\; \end{equation} and \begin{equation} \dot
x_1=-12\wp_{23},\;\dot x_2=-12\wp_{13},\;\dot x_3=-12\wp_{12},
\end{equation} The entries $U$ and $Q$ to the matrices $L$ and $M$ have
the form \begin{eqnarray} U(z)&=&4(z-X_1)(z-X_2)(z-X_3),\label{uu3}\\
Q(z)&=&4z^3-9g_2z+8X_1X_2X_3+27g_3,\label{qq3} \end{eqnarray} where, in
this case, $X_i=3\wp_{jk}$. The curve ${\rm det}(L(z)-yI)=0$ has the
form \begin{equation}
y^2={1\over16}(4z^2-3g_2)^2(z^2-3g_2)(4z^3-9g_2z+27g_3).\label{sepu3}
\end{equation}

The Lax representations allow to construct the linear $r$-matrix
algebra of the form (\ref{linal}) which we will discuss elsewhere.

\section{Elliptic Potentials from the Theta Functional Point of View}

Let us fix the homology basis $(A,B)=(A_1,\ldots,A_g;B_1,\ldots,B_g)$
on the curve (\ref{curve}) and a canonically conjugated basis of
holomorphic differentials ${\bf v}=(v_1,\ldots,v_g)$ in such a  way
that the Riemann matrix has the form \[ \left(\oint_{A_1}{\bf
v},\ldots,\oint_{A_{g}}{\bf v}; \oint_{B_1}{\bf
v},\ldots,\oint_{B_{g}}{\bf v}\right)= ({\bf 1}_{g};\tau).\] with the
matrix $\tau$ belonging to Siegel upper half space ${\cal S}_g$ of
degree $g$.  Let us denote by $ {\bf A}(Q) =\int^{Q}_{\infty }{\bf v}$
the Abel map $C_2\rightarrow {J(C_g)}$, where $J(C_g)$ is the Jacobian
of the curve $C_g$.

Let us determine the Riemann theta function $\theta [\varepsilon]({\bf
z}|{\tau})$  on ${\bf C}^g\times {\cal S}_g$  with  the characteristics
\[ [\varepsilon]   = \left[ \begin{array}{l} \varepsilon' \\
\varepsilon''\end{array} \right] = \left[
\begin{array}{lll}\varepsilon_1'&\ldots&\varepsilon_g'\\
\varepsilon_1''&\ldots&\varepsilon_g'' \end{array}\right], \]
 by the formula \begin{eqnarray} &&\theta [\varepsilon]({\bf z}|{\tau})
\label{theta}\\&&=\sum_{ \displaystyle{{\bf m} \in {\Bbb Z}^g}} \exp
\pi i \{ \langle ({\bf m} + {\varepsilon'\over2}){\tau}, ({\bf m}
+{\varepsilon'\over2}) \rangle+ 2\langle ({\bf m} +
{\varepsilon'\over2}), {\bf z} + {\varepsilon''\over2} \rangle\},
\nonumber\end{eqnarray} where $\langle \cdot , \cdot\rangle$ means the
Euclidean scalar product. For integer characteristics we have
\begin{eqnarray}
\theta\left[{}_{\varepsilon^{\prime}}^{\varepsilon^{\prime\prime}}\right]
({\bf z}|\tau)= {\rm exp}\,2\pi\,\left[
{1\over2}{{}^t\varepsilon^{\prime}\over2}\tau{\varepsilon^{\prime}\over2}
+{\varepsilon^{\prime}{\bf z}\over2}+
{1\over2}{{}^t\varepsilon^{\prime}\over2}{\varepsilon^{\prime\prime}\over2}
\right]\theta\left({\bf z}+I{\varepsilon^{\prime\prime}\over2}+ \tau
{\varepsilon^{\prime}\over2}|\tau\right) .\label{shift} \end{eqnarray}

If  $\varepsilon_i',\varepsilon_j''$   are equal to $0$ or  $1$, the
characteristics $[\varepsilon]$   are the characteristics of the
half--periods. The theta function  (\ref{theta})  is odd or even  if
$[\varepsilon]$    is a half-period  characteristic, and  we call the
corresponding $[\varepsilon]$  odd or even.

The function (\ref{theta})  satisfies  the  two  sets of
 functional equations  \cite{mu83}, the {\it transformational property}
\begin{eqnarray} &&\theta[\varepsilon]({\bf z}+{\bf n}''+{\bf
n}'{\tau}|{\tau}) \nonumber\\&&= \exp \pi i\Bigl[ -\langle {\bf
n}'{\tau},{\bf n}'\rangle - 2\langle {\bf n}'' , {\bf z}\rangle
  +\langle \varepsilon',{\bf n}'\rangle - \langle\varepsilon'',{\bf n}'
\rangle\Bigr]\,\theta[\varepsilon]({\bf z}|{\tau}) \label{trans}
\end{eqnarray} where ${\bf n', n''} \in {\Bbb Z}^g$ and the {\it
modular property}, which describes the transformation of the theta
function under the action of the group $Sp_{2g}({\Bbb Z})$.

The almost-periodic function $u(x)$ is called a {\it finite-gap
potential} if the spectrum of the Schr\"odinger operator
$H=-\partial_x^2+u(x)$ is a union of the finite set of segments of a
Lebesque (double absolutely continuous) spectrum. Let us formulate the
Its-Matveev theorem \cite{im75}

\begin{theorem} (Its--Matveev theorem).  The potential $u(x)$ and the
eigenfunction $\Psi(Q,x)$ of the Schr\"odinger operator
$H=-\partial_x^2+u(x)$ associated with the $g$--gap Lebesque spectrum
$\Sigma=[z_1z_2]\cup[z_3,z_4]\cup\ldots\cup[z_{2g+1},\infty]$, are
expressed by the formula \begin{eqnarray}
u(x)&=&-2{\partial^2\over\partial x^2}{\rm ln} \theta(i{\bf U}x+{\bf
A}(Q)-{\bf A}({\cal D})|{ \tau})+C,\label{pot}\\ \Psi(Q,x)&=&
{\theta(i{\bf U}x+{\bf A}(Q)+{\bf A}({\cal D})|{ \tau})\over
\theta(i{\bf U}x+{\bf A}({\cal D})|{\bf \tau})} {\rm exp}\left(ix
\int^Q_{\infty}\Omega\right).\label{wave} \end{eqnarray} Here $Q$ is
the point of a hyperelliptic Riemann surface $C_g=(w,z)$ defined by the
equation $w^2  = \prod_{j=1}^{2g+1} (z - z(Q_j)),
 \, z(Q_j) = z_j \in {\bf C} ,\, z_i \neq z_j,$ a nonsingular
hyperelliptic curve of genus  $g$, realized by means of the function
$z$  as a 2--sheeted covering of the Riemann sphere with  branching
points at $Q_1,\ldots,Q_{2g+1}$, $Q_{2g+2}$, $z(Q_{2g+2})  =  \infty$.
${\bf v}$ is the vector of normalized holomorphic differentials, $\tau$
is the matrix of their periods, ${\bf A}(Q)=\int_{\infty}^Q{\bf v}$,
$\Omega$ is a normalized Abelian differential of the second kind which
has a second-order pole at the infinity with the principal part
$\zeta^{-2}d\zeta$, where $\zeta$ is a local variable, ${\bf U}$ is the
vector of periods of the differential $\Omega$, ${\cal D}$ is
nonspecial divisor, ${\bf K}$ is the vector of Riemann constants.
\end{theorem}

The components $U_i,\,V_i,\, i = 1,\ldots,g$ of the {\it winding
vectors} {\bf U}, {\bf V} in (\ref{pot},\ref{wave}) are expressed in
terms of the normalizing constants $c_{ij}$ of the holomorphic
differentials and projections of the branching points
$z_1,\ldots,z_{2g+1}$ by the formulae \begin{eqnarray} U_j=-2i c_{1j},
\quad V_j=-i\left(2c_{2j}+ c_{1j}\sum_{k=1}^{2g+1}z_k \right), \quad
j=1,\ldots,g.  \label{wind} \end{eqnarray}

Further we shall restrict ourselves the case of  genus two curves.

Let us give the theta functional construction of the two-gap elliptic
potentials.  Following  Sect. 2, we describe such points $\tau \in
{\cal S}_{2}$, for which the function (\ref{pot}) is elliptic.  For
this purpose we consider the {\it Humbert surface} $H_{\Delta },
\Delta  = N^{2}$, i.e.\  the variety \begin{eqnarray}
 H_{\Delta} &=&\left\{\tau\alpha \tau_{11} + \beta \tau_{12} + \gamma
\tau_{22} + \delta (\tau^{2}_{12} - \tau_{11}\tau_{22}) +\varepsilon
=0,\right.\nonumber\\&& \left.  \alpha,\beta , \gamma , \delta ,
\epsilon  \in  {\Bbb Z}, \quad \Delta = \beta^{2} - 4(\alpha \gamma  +
\varepsilon \delta)\right\}.\label{hs} \end{eqnarray} The quantity
$\Delta$  is an invariant with respect to the action of the group
$Sp_4(\bf Z)$ \cite{kr03}.  The following theorem summarizes the
Weierstrass reduction theory for the case of genus two algebraic curves
(see e.g. \cite{kr03},\cite{bbeim94}).

\begin{theorem} (Reduction theorem) Let $C_{2}$  and $C_{1}$   be  the
curves of genus two and one, which are equipped  by  the homology basis
$(A_{1},A_{2};B_{1},B_{2})$ and $(A,B)$. The curve $C_2$  is an
$N$-sheeted covering of the torus $C_{1}$ if and only if the moduli of
$C_{2}$  belong  to the Humbert surface  with $\Delta  = N^{2}$ and the
integer numbers $\alpha, \beta, \gamma, \delta, \epsilon$, being
expressed in terms of of the  elements  of  the  integer matrix $M$,
mapping the basis $(A_1,A_2,B_1,B_2)$ into the basis $(A,B)$ \[
M\left(\begin{array}{l}A_1\\ A_2\\ B_1\\ B_2\end{array}\right) =
\left(\begin{array}{l}A\\ B\end{array}\right) ,\quad
M=\left(\begin{array}{cc}m_{11} & m_{12}\\ m_{21} &m_{22}\\ m_{31} &
m_{32}\\m_{41} & m_{42}\end{array}\right), \] are given by the
following formulae \begin{eqnarray} \alpha &=&m_{12} m_{41}-m_{12}
m_{42}{,}\quad{\gamma } =m_{21} m_{32}{-} m_{31} m_{22},\nonumber\\
\delta&=&m_{12} m_{21}-m_{11} m_{22}{,}\quad{\epsilon }
=m_{31}m_{42}{-} m_{41} m_{32},\nonumber\\ \beta&=&m_{11} m_{32}-m_{31}
m_{12}-(m_{21}m_{42}-m_{41} m_{22}).  \label{mm} \end{eqnarray}

Moreover there exists an element $\sigma \in  Sp_{4}({\Bbb Z})$ and a
point $\tau \in  {\cal S}_2$ such that \begin{equation} \sigma \circ
\tau = \left(\begin{array}{ll}\tau_{11}&{1\over N}\\ {1\over N
}&\tau_{22}\end{array}\right).\label{wetau} \end{equation}
\end{theorem}

Under the conditions of the reduction theorem, the two-dimensional theta
function is reduced with the help of the addition theorem for theta
functions of $N$-th order (see e.g. \cite{ko79}) to the finite sum of
products of Jacobian theta functions with the moduli $N\tau_{11}$ and
$N\tau_{22}$.

Below we apply the Weierstrass reduction theory to describe all elliptic
genus two potentials.

\begin{lemma} The function \begin{equation} f(x)=-2\partial_x^2\,{\rm
ln}\theta\left({x\over 2\omega}+\alpha,\beta\big|{1\over
N}\left(\begin{array}{ll}{\omega'\over
\omega}&1\\1&\widetilde\tau\end{array}\right)\right) \label{f}
\end{equation} with arbitrary $(\alpha,\beta)\in J(C_2)$, ${\rm
Im}\,\omega'/\omega={\rm Im}\, \tau>0$ is an $N$-th order elliptic
function with primitive periods $2\omega$, $2\omega'$ and can be
represented in the form \begin{equation}
f(x)=2\sum_{j=1}^n\wp(x-x_j)+6\sum_{k=1}^m\wp(x-x_k),\quad n+3m=N
\label{gap2} \end{equation} with ${x_j}$ belonging to the locus ${\cal
L}_N$ or its closure.  \end{lemma}

{\bf Proof} It follows from the transformational properties of theta
functions that the function (\ref{f}) is a doubly periodical function
on the torus $C_1$ with the primitive periods $2\omega$, $2\omega'$ and
$\tau=\omega'/\omega$. Let us calculate the number of poles of the
(\ref{f}). To do this we consider the function \begin{equation} g(x)=
{\theta\left({x\over 2\omega}+\alpha,\beta|{1\over N}\left(
\begin{array}{ll}{\omega'\over
\omega}&1\\1&\tau'\end{array}\right)\right) \over
\vartheta_3\left({Nx\over 2\omega}+N\alpha|\tau\right)}, \label{g}
\end{equation} where $\vartheta_3$ is a Jacobi theta function. The
function $f(x)$ is meromorphic on the torus $C_1$ as follows from the
transformational properties of the theta function. Further, the
denominator (\ref{g}) has exactly $N$ zeros in $C_1$:  \[ x={2k+1\over
N}\omega'+\omega-2\alpha\omega,\quad k=0,1,\ldots,N-1.  \] Therefore,
according to the Abel theorem, the numerator has exactly $N$ zeros,
these are, $x_1,\ldots,x_N$.  To prove that the function (\ref{f}) can
be written in the form (\ref{gap2}) we note that the function (\ref{f})
is a two--gap potential and therefore the corresponding wave function
can have a pole of no more than second order. Using the Schr\"odinger
equation we find the coefficients 2 and 6 in the decomposition
(\ref{gap2}). The proof that the points $x_1,\ldots,x_N$ belong to the
locus ${\cal L}_N$ is carried out by substituting of the ansatz
(\ref{gap2}) into the KdV equation and equating the principal parts of
poles to zero.

\begin{theorem} (Main theorem) The two--gap potential as defined by
formula (\ref{pot})  is an elliptic function of the $N$-th order if and
only if\\ $1) \quad C_{2}$  covers a torus $C_1$ $N$-sheetedly,\\ $
2)\quad U_{1} U_{2} = 0.$ \end{theorem}

{\bf Proof}. Sufficiency. Suppose the conditions of the theorem are
fulfilled and for definiteness $U_1 = 0.$  Then the function
(\ref{pot}) is an elliptic function of order $N$  according to the
Lemma.

Necessity. Let the evolution of an  integrable  dynamic  system with
two  degree  of  freedom  be  described  in  terms  of  elliptic
functions. Alternatively this evolution is expressed in terms of theta
functions defined on the Jacobian $J(C_{2})$. If $\omega ,
\omega^{\prime} , {\rm Im}\,\omega /\omega^{\prime } > 0$, are the
primitive periods of elliptic  function, then  the  following
identities are valid due to the transformation property of the theta
function (\ref{trans}).  \begin{eqnarray} 2{U_{1}}{\omega }
&=&r({n}{+}{p^\prime }{\tau_{11}}{+}{q^\prime }{\tau_{12}}),\quad
2{\omega }_1{\omega }=r({m}{+}{p^\prime }{\tau}_{12}{+}{q^\prime
}{\tau}_{22})\nonumber\\ 2{U_{2}}{\omega^{\prime} } &=&s({n^{\prime}
+}{p}{\tau}_{11}{+}{q}{\tau_{12}}),\quad 2{U_{2}}{\omega^{\prime} }
=s({m^{\prime} +}{p}{\tau_{12}}{+}{q}{\tau_{22}})\label{elcon}
\end{eqnarray} \noindent where $n, m, n^{\prime} , m^{\prime} , p, q,
p^{\prime} , q^{\prime}  \in  {\Bbb Z}$.  Eliminating
$U_{i}\omega^{\prime} $, $U_{i}\omega, i= 1, 2$  from (\ref{elcon}) we
find that $\tau$   belongs  to the Humbert surface $H_{\Delta }$, with
\begin{eqnarray} \alpha  &=& m^{\prime} p^{\prime} - mp,\,\delta
=pq^{\prime }-p^{\prime} q,\nonumber\\ \gamma  &=&nq - n^{\prime}
q^{\prime},\,\epsilon =nm^{\prime }-mn^{\prime }.\nonumber\\ \beta
&=&np-m^{\prime} q^{\prime }-mp-n^{\prime} p^{\prime },\label{abc}
\end{eqnarray}

Calculating the invariant $\Delta $, defined in (\ref{hs}), we find
that $\Delta  = N^{2} , N = n p + mq - m^\prime q^{\prime}  -
n^{\prime} p^{\prime} $.  Therefore  the assumption  of  the theorem
leads to the conclusion that $C_{2}$ covers a torus  $N$-sheetedly.
But in this case we can define a matrix $M$ which maps the homology
basis on $C_{2}$ to the homology  basis  on $C_{1}$.  Taking into the
account (\ref{mm}) we find \begin{equation} {M}{=}
\left(\begin{array}{ll}p&-p^{\prime} \\ q&-q^ {\prime} \\ -n{^\prime}
&n\\ -m^{\prime}
&m\end{array}\right){,}\quad{m}{p}{+}{m}{q}{-}{m^{\prime} q^{\prime}
}{-}{n^{\prime} p{^\prime} }={N}\label{elcon2} \end{equation} According
to the reduction theorem there exists a transformation $\rho$ which
maps the matrix $\tau$ to the form  (\ref{wetau}). Therefore we  have
in  the  new homology basis \begin{eqnarray} p&=&N,\quad q=0,\quad
n=1,\quad m=0,\nonumber\\ p^{\prime} & =& 0,\quad q^{\prime}  = 0,
\quad n^{\prime}  = 0,\quad m^{\prime} = -1.\label{elcon1}
\end{eqnarray}
{}From (\ref{elcon},\ref{elcon2}) we conclude, that
\begin{equation} 2{U_{1}\omega } =r,\quad {U_{1}}{\omega^{\prime }}
=s{N}{\tau_{11},}\quad {U_{2}}={0}\label{3.15} \end{equation} and the
theorem is proved.

It follows from the conditions of the theorem that elliptic potentials
are singled out from finite-gap potentials by some subvariety in the
Humbert variety. We shall call this variety  $E_{\Delta}$-{\it
variety}, $E_{\Delta}\in H_{\Delta}$.

Let us derive the two--gap Lam\'e and Treibich--Verdier potentials from
the reduction technique of finite gap potentials to elliptic potential
developed above.

\begin{prop} (Proposition on the Treibich-Verdier potentials) The only
two--gap primitive Lam\'e and Treibich--Verdier potentials are the
three first elliptic  two--gap elliptic potentials  from   Table 1.
\end{prop}

{\bf Proof} Let us consider the elliptic potential \begin{equation}
u(x)=-2\partial_x^2\,{\rm ln}\, \theta[\delta]\left({x\over
2\omega},0|\left(\begin{array}{ll} {1\over N}{\omega'\over
\omega}&{1\over N} \\{1\over N}& {1\over N}{\widetilde\omega'\over
\widetilde\omega} \end{array}\right)\right),\; U_2=0, \label{elpot}
\end{equation} with $[\delta]$ running through all the six odd
characteristics.  Let consider the function \[
\Theta(x)=\theta[\delta]\left({x\over2\omega}|\left(\begin{array}{ll}
{1\over N}{\omega'\over \omega}&{1\over N}\\{1\over N}& {1\over
N}{\widetilde\omega'\over \widetilde\omega}\end{array}\right)\right).
\] At $x=0$ this is a theta constant with the characteristic
$[\delta]$. One can calculate using (\ref{shift}) that at $x=\omega $
the characteristics becomes $[\delta]+[{}^0_1{}_0^0]$, at
$x=\omega^{\prime}$ the charteristic $[\delta]$ turns into the
characteristic $[\delta]+[{}^N_0{}^N_1]$ and at $x=\omega+\omega'$ it
is $[\delta]+[{}^N_1{}^N_1]$. Let us denote by $(n_0,n_1,n_2,n_3)$ the
coefficients in the decomposition
$u_N=n_0\wp(x)+n_1\wp(x+\omega)+n_2\wp(x+\omega+\omega')+n_3\wp(\omega')$
with  $\sum_{k=0}^3 n_k=N\label{n}$ and $n_i= 2$ or 6.  Let the
characteristic $[\delta ]$ runs through all the odd characteristics.
Then for odd $N$ we have
$$\begin{array} {lllll}
x=0&x=\omega&x=\omega'&x=\omega'+\omega'&(n_0,n_1,n_2,n_3)\\
\left[{}_1^1{}_0^0\right]&\left[{}_0^1{}_0^0\right]&
\left[{}_1^0{}_1^1\right]&\left[{}_0^0{}_1^1\right]&
(n_0,0,n_2,n_3)\\
\left[{}_1^1{}_1^0\right]&\left[{}_0^1{}_1^0\right]&
\left[{}_1^0{}_0^1\right]&\left[{}_0^0{}_0^1\right]&
(n_0,0,0,0)\\
\left[{}_1^1{}_0^1\right]&\left[{}_0^1{}_1^0\right]&
\left[{}_1^0{}_1^0\right]&\left[{}_0^0{}_1^0\right]&
(n_0,0,0,0)\\
\left[{}_0^0{}_1^1\right]&\left[{}_1^0{}_1^1\right]&
\left[{}_0^1{}_0^0\right]&\left[{}_1^1{}_0^0\right]&
(n_0,n_1,0,n_3)\\
\left[{}_1^0{}_1^1\right]&\left[{}_0^0{}_1^1\right]&
\left[{}_1^1{}_0^0\right]&\left[{}_0^1{}_0^0\right]&
(n_0,n_1,n_2,0)\\
\left[{}_0^1{}_1^1\right]&\left[{}_1^1{}_1^1\right]&
\left[{}_0^0{}_0^0\right]&\left[{}_1^0{}_0^0\right]&
(n_0,0,0,0) \end{array}$$
We see that the only possibilities are $u(x)=6\wp(x)$ or
$2\wp(x+\omega_1)+2\wp(x+\omega_2) +2\wp(x+\omega_3)$ and
$u(x)=6u(x)+2\wp(x+\omega_i)+2\wp(x+\omega_k)$.

For even $N$ we have
$$\begin{array} {lllll}
x=0&x=\omega&x=\omega'&x=\omega'+\omega'&(n_0,n_1,n_2,n_3)\\
\left[{}_1^1{}_0^0\right]&\left[{}_0^1{}_0^0\right]&
\left[{}_1^1{}_1^0\right]&\left[{}_0^1{}_1^0\right]& (n_0,0,n_2,0)\\
\left[{}_1^1{}_1^0\right]&\left[{}_0^1{}_1^0\right]&
\left[{}_1^1{}_0^0\right]&\left[{}_0^1{}_0^0\right]& (n_0,0,n_2,0)\\
\left[{}_1^1{}_0^1\right]&\left[{}_0^1{}_0^1\right]&
\left[{}_1^1{}_1^1\right]&\left[{}_0^1{}_1^1\right]& (n_0,0,0,n_3)\\
\left[{}_0^0{}_1^1\right]&\left[{}_1^0{}_1^1\right]&
\left[{}_0^0{}_0^1\right]&\left[{}_1^0{}_0^1\right]& (n_0,n_1,0,0)\\
\left[{}_1^0{}_1^1\right]&\left[{}_0^0{}_1^1\right]&
\left[{}_1^0{}_0^1\right]&\left[{}_0^0{}_0^1\right]& (n_0,n_1,0,0)\\
\left[{}_0^1{}_1^1\right]&\left[{}_1^1{}_1^1\right]&
\left[{}_0^1{}_0^1\right]&\left[{}_1^1{}_0^1\right]& (n_0,0,0,n_3)
\end{array}$$ We see that the only possibilities in this case are
$u(x)=6\wp(x)+2\wp(x+\omega_i)$. The proposition is proved.

\subsection{Elliptic Subvarieties of Humbert Surfaces}

The components of the Humbert surface are described in terms of the
vanishing of some modular forms \cite{kr03}, more generally, the
Humbert surface $H_{\Delta}$ is described by some ideal in the ring of
modular forms \cite{ge82}. Therefore it is natural to describe elliptic
subsurfaces $E_{\Delta}$ of $H_{\Delta}$, $\Delta=N^2$ in terms of the
vanishing of some theta constants.

\begin{prop} (Proposition on the elliptic points) Let the nonsingular
curve associated with the two-gap potential cover a torus
$N$-sheetedly. Let us fix such a homology basis that the matrix $\tau $
has the form (\ref{wetau}) and belongs to the component $H_{\Delta}$.
Then elliptic points in $H_{\Delta}$ are separated by the condition
\begin{eqnarray}
\theta_i[\delta]\left(0\vert\left(\begin{array}{ll}\tau & {1\over
N}\\{1\over N}&\widetilde\tau\end{array}\right)\right) =
0,\label{ellip} \end{eqnarray} where  $[\delta]$ runs through all the
six odd characteristics and $i=1$ or 2 \end{prop}

{\bf Proof} It follows immediately from the {\it Rosenhain formulae for
the normalising constants of the holomorphic
differentials\/}\footnote{These formulae are a consequence of the
important {\it Rosenhain derivative formulae\/} given in  Appendix B}
and the expression (\ref{wind}) for the winding vectors. Assume that
the curve $C_2$  has the form
$w^2=z(z-1)(z-\lambda_1)(z-\lambda_2)(z-\lambda_3)$, then the
normalizing constants of the holomorphic differentials
$v_i=(c_{i1}z+c_{i2})dz/w, i=1,2$  have the form \cite{kw15}
\begin{equation} c_{i1}=-{\theta_1[\varepsilon_i]\over
2\pi^2\theta[\delta_{i1}]\theta[\delta_{i2}]
\theta[\delta_{i3}]},\label{rosen} \end{equation} where
$[\varepsilon_i]$ is an odd theta constant and $[\delta_i],i=1,2,3$ are
such even theta constants that
$[\varepsilon_i]=[\delta_{i1}]+[\delta_{i2}]+[\delta_{i3}]$.

Let us give a few examples for the condition (\ref{ellip}). The
simplest ones are at $N=2^p, p=1,\ldots$ because to simplify we can
apply  the the  addition theorem for the theta functions of the second
order (see e.g. \cite{mu83}) \begin{eqnarray}
&&\theta[\varepsilon]({\bf x}|\tau)\theta[\delta]({\bf
y}|\tau)\label{add}\\ &=&\sum_{\rho}
\theta\left[\begin{array}{c}{1\over2}(\varepsilon^{\prime}+\delta^{\prime})
+\rho
\\\varepsilon^{\prime\prime}+\delta^{\prime\prime}\end{array}\right]
({\bf x}+{\bf y}\big|2\tau)
\theta\left[\begin{array}{c}{1\over2}(\varepsilon^{\prime}-\delta^{\prime})
+\rho
\\\varepsilon^{\prime\prime}-\delta^{\prime\prime}\end{array}\right]({\bf
x}-{\bf y}\big|2\tau) ,\nonumber \end{eqnarray} where the summation
runs over $\rho =(0,0),(0,1),(1,0),(1,1)$.  The particular cases of
(\ref{add}) which are necessary for the calculations are given in
Appendix B. We also use below the formula \begin{eqnarray}
&&\theta\left[\begin{array}{cc}\varepsilon_1^{\prime}&\varepsilon_2^{\prime}
\\\varepsilon_1^{\prime\prime}&\varepsilon_2^{\prime\prime}\end{array}\right]
\left({\bf
z}\big|\left(\begin{array}{cc}\tau&1\\1&\widetilde\tau\end{array}
\right)\right)\nonumber\\&&={\rm e}^{-{1\over2}\pi i
\varepsilon_1^{\prime}\varepsilon_2^{\prime}}
\theta\left[\begin{array}{cc}\varepsilon_1^{\prime}&\varepsilon_2^{\prime}
\\\varepsilon_1^{\prime\prime}+\varepsilon_2^{\prime}&\varepsilon_2^{\prime\prime}+\varepsilon_1^{\prime}\end{array}\right]
\left({\bf
z}\big|\left(\begin{array}{cc}\tau&0\\0&\widetilde\tau\end{array}
\right)\right)\label{red} \end{eqnarray}

For example, the condition (\ref{ellip})  for $N=2$ is
$\vartheta_1^{\prime}\vartheta_3\widetilde \vartheta_4^2=0$, where
$\vartheta_i=\vartheta_i(0|2\tau)$,
 $\widetilde\vartheta_i=\vartheta_i(0|2\widetilde\tau)$.  This
condition is not satisfied for nonsingular tori. Therefore elliptic
potentials of the form $2\wp(x-x_1)+2\wp(x-x_2)$ do not exist.

{\bf Example: N=4} For $N=4$ the condition (\ref{ellip}) written for
the characteristic $[{}^1_0{}^1_1]$ reads \begin{equation}
{\sqrt{2}\vartheta^2_2\widetilde\vartheta_3\over\vartheta_3}+
\sqrt{\vartheta_3^2\widetilde\vartheta_3^2+
\vartheta_2^2\widetilde\vartheta_4^2 -
\vartheta_4^2\widetilde\vartheta_2^2}=0.\label{zz} \end{equation} To
obtain (\ref{zz}) we  used (\ref{add}) twice.

The condition (\ref{zz}) rewritten in terms of the Jacobi moduli
$k={\vartheta_2^2/\vartheta_3^2}$, $\widetilde k={\widetilde
\vartheta_2^2/\widetilde \vartheta_3^2}$, where
$\vartheta_i=\vartheta_i(0|4\tau)$,
$\widetilde\vartheta_i=\vartheta_i(0|4\widetilde\tau)$,
coincides with those given in  Table 5.

Let us derive the potential $u_4$ by  direct computation. We have
according to  Theorems 1 and 2
\begin{eqnarray}
u_4(x)&=&\partial^2_x\,{\rm
ln}\,\Theta(x)+C,\label{potu4} \\
\Theta(x)&=&\theta[\delta]
\left({x\over 2\omega},0\big| \left(\begin{array}{ll}
{\tau}&{1/4}\\{1/4}&{\widetilde \tau}\end{array}\right)\right).
\nonumber
\end{eqnarray}
Let us consider the definite case $[\delta]=\left[{}_0^1{}_1^1\right]$
Applying  (\ref{add}) twice we have \begin{eqnarray} \Theta(x)&=&
8{\left( \widehat{\widehat{\theta}}\left[{}_0^1{}_0^0\right]({\bf z})
\widehat{\widehat{\theta}}\left[{}_0^0{}_0^0\right]({\bf z})
-\widehat{\widehat{\theta}}\left[{}_0^0{}_0^1\right]({\bf z})
\widehat{\widehat{\theta}}\left[{}_0^1{}_0^1\right]({\bf z})
\right)\over \widehat\theta\left[{}_1^0{}_0^1\right]
\widehat\theta\left[{}_0^0{}_0^1\right] }\nonumber\\ &\times&
\left(\widehat{\widehat{\theta}}\left[{}_0^0{}_1^1\right]({\bf z})
\widehat{\widehat{\theta}}\left[{}_0^0{}_1^0\right]({\bf z})
+\widehat{\widehat{\theta}}\left[{}_0^1{}_1^0\right]({\bf z})
\widehat{\widehat{\theta}}\left[{}_0^1{}_1^1\right]({\bf z})
\right)\nonumber\\
&+&4{\left(\widehat{\widehat{\theta}}\left[{}_0^1{}_1^0\right]({\bf z})
\widehat{\widehat{\theta}}\left[{}_0^0{}_1^1\right]({\bf z})
+\widehat{\widehat{\theta}}\left[{}_0^1{}_1^1\right]({\bf z})
\widehat{\widehat{\theta}}\left[{}_0^0{}_1^0\right]({\bf z})
\right)\over \widehat\theta\left[{}_0^1{}_0^1\right]
\widehat\theta\left[{}_0^0{}_1^0\right]}\nonumber\\ &\times&
\left(\widehat{\widehat{\theta}}^2\left[{}_0^0{}_0^0\right]({\bf z})
-\widehat{\widehat{\theta}}^2\left[{}_0^1{}_0^1\right]({\bf z})
+\widehat{\widehat{\theta}}^2\left[{}_0^1{}_0^0\right]({\bf z})
-\widehat{\widehat{\theta}}^2\left[{}_0^0{}_0^1\right]({\bf z})
\right), \label{xx} \end{eqnarray} where we denote
$\widehat\theta[\epsilon]=\theta[\epsilon](0|2\tau),
\;\widehat{\widehat{\theta}}(z)=\theta[\epsilon]({\bf z} |4\tau)$ and
${\bf z}=(x/2\omega,0)$.

Using  (\ref{red}), we rewrite (\ref{xx}) in the form \begin{eqnarray}
\Theta(x)&=&4i{ \vartheta_1({x\over 2\omega})\vartheta_3({x\over
2\omega})\widetilde\vartheta_2\vartheta_4\over\sqrt{\vartheta_2\vartheta_4}
\widetilde\vartheta_3\widetilde\vartheta_4}\nonumber\\ &\times&\left\{
{\sqrt{2}\vartheta^2_2({x\over 2\omega})\widetilde\vartheta_3\over
\vartheta_3}+{ \vartheta_3^2({x\over2\omega})\widetilde\vartheta_3^2+
\vartheta_2^2({x\over2\omega})\widetilde\vartheta_4^2 -
\vartheta_4^2({x\over2\omega})\widetilde\vartheta_2^2
\over\sqrt{\vartheta_3^2\widetilde\vartheta_3^2+
\vartheta_2^2\widetilde\vartheta_4^2
-\vartheta_4^2\widetilde\vartheta_2^2}}\right\}.  \end{eqnarray}

By the condition (\ref{zz}) and addition formulae for Jacobi theta
functions \cite{Ba55}, one can prove that $\Theta(x)$ is proportional
to $\vartheta_1^3({x\over 2\omega})\vartheta_2({x\over 2\omega})$ and
therefore the potential $u_4$ has the form given in  Table 1.

{\bf Further examples.} Let us consider the function \begin{equation}
u(x)=\partial^2_x\,{\rm ln}\,\theta[\delta] \left({x\over 2\omega},0\big|
\left(\begin{array}{ll}
{\tau}&{1/2^p}\\{1/2^p}&{\widetilde\tau}\end{array}\right)\right),
\end{equation} $p>2$ and the moduli $\tau$ and $\tilde\tau$ are
connected by the condition \begin{eqnarray}
\theta_i[\delta]\left(0\vert\left(\begin{array}{ll}\tau & {1\over
2^p}\\{1\over 2^p}&\widetilde\tau\end{array}\right)\right) =
0.\label{ellip3} \end{eqnarray}

One can show (see the formulae for theta constants of the $2^p$-sheeted
cover) that the (\ref{ellip3}) is valid and $E_{2^p}$ is not empty for
$p=2,3,..$ In particular denoting
$X=i{\vartheta_2(0;2^p\tau)/\vartheta_3(0;2^p\tau)}$,
$Y={\vartheta_2(0;2^p\tilde\tau)/\vartheta_3(0;2^p\tilde\tau)}$ we plot
below two varieties $E_{2^p}$ for $p=2$ and $p=3$ respecively, in the
coordinates $X,Y$.

The curve given on the plot \footnote{The plot can be sent if
required from addresses, e-mail: chris@cara.ma.hw.ac.uk  or
vze@cara.ma.hw.ac.uk)}
corresponds to a family of elliptic potentials.  We emphasise that the
potential $u_8$ is a new elliptic potential connected with an
eight-sheeted cover of the torus.  It differs from the $u_8$
Treibich-Verdier potential which is not primitive and can be obtain
from the Treibich-Verdier potential $u_4$ by the Gauss transform
(\ref{gauss}) of the moduli of one of the tori.

We can summarise all the discussion by the following statement:
\begin{conj} There exist infinitely many primitive elliptic potentials
$u_N(x)$ of genus two at $N \in {\Bbb N}$. Therefore the two gap locus
of Calogero-Moser system has infinitely many components.  \end{conj}

The hypothesis is valid at least for $N=4,5,8,$ as   shown in this
paper.  To prove it it is sufficient to find solutions for
(\ref{ellip}) for a countable number of $N$.

\section*{Acknowledgements} The authors are grateful to E D Belokolos
for many suggestions for reduction techniques for finite-gap potentials
and  to H W Braden, A P Fordy, V B Kuznetsov, E K Sklyanin and T.Suzuki
for valuable discussions.  VZE would like to acknowledge the support of
the Royal Society.  \newpage \appendix

\section{Two-gap Lam\'e and Treibich-Verdier Potentials} We give in
this Appendix the complete description of two primitive
Treibich-Verdier potential which includes the explicit formulae for the
covers, link between moduli of the tori, wave functions and Lam\'e
polynomials. We also give for the complicity the analogous description
of the two-gap Lam\'e potential, which is known.  We note that some of
this results concerning Treibich-Verdier potentials were first given in
\cite{sm93} and \cite{ek93}.

\begin{table}[htpb] \caption{The spectral curves}
\begin{tabular}{||l|l|l||}\hline $u_N$ &The spectral curve $C_2=(w,z)$&
The coordinate $\lambda$
 \\{}&{}& in terms of $w,z$\\ \hline
$u_3$&$\rule[-2mm]{0mm}{7mm}w^2=-(z^2-3g_2)\prod_{i=1}^3(z-3e_i)$&{}\\
\cline{2-2}
   {}&$\rule[-1mm]{0mm}{6mm}\lambda^3-3\lambda\wp+\wp^{\prime}=0$&$
\rule[-3mm]{0mm}{7mm}\lambda=\frac{2w}{3z^2-3g_2}$\\ \hline
$u_4$&$\rule[-2mm]{0mm}{7mm}w^2=(z+6e_i)\prod_{k=1}^4(z-z_k(i))
,i=1,2,3$&{}\\
   {}&$z_{1,2}(i)=e_j+2e_i\pm\sqrt{(e_i-e_j)(2e_j+7e_i)}$&{}\\
{}&$z_{3,4}(i)=e_k+2e_i\pm\sqrt{(e_i-e_k)(2e_k+7e_i)}$&
{$\lambda=\frac{3w}{2(6e_j+z)(15e_j-2z)}$}\\ \cline{2-2}
   {}&$\rule[-2mm]{0mm}{7mm}\lambda^4-3(2\wp-e_i)\lambda^2
   +4\lambda\wp^{\prime}-3(\wp-e_j)(\wp-e_k)=0,$&{}\\ {}&$i\neq j\neq k
   = 1,2,3$&{}\\ \hline
$u_5$&$\rule[-2mm]{0mm}{7mm}w^2=\prod_{i=1}^{i=5}(z-z_i(j)), \quad
j=1,2,3${}&\\ {}&$z_4(j)=6e_k-3e_i,\;z_5(j)=6e_i-3e_k$&{}\\
{}&$\prod_{i=1}^3(z-z_i(j))$&$\lambda=\frac{-4w}
{5z^2+6ze_j+261e_j^2-108g_2}$\\
{}&$\rule[-2mm]{0mm}{6mm}=z^3-3z^2e_j+(51e_j^2-20g_2)z
-369e_j^3+132e_jg_2$&\\ \cline{2-2}
{}&$\rule[-2mm]{0mm}{7mm}\lambda^5-2(e_i+5\wp)\lambda^3+10\wp
\lambda^2+3(e_i+5\wp)(e_i-\wp)\lambda$&\\
{}&$\rule[-2mm]{0mm}{7mm}-2\wp^{\prime}(e_i-\wp)=0${}&\\\hline
\end{tabular} \end{table}

\begin{table}[htpb] \caption{The first cover}
\begin{tabular}{||l|l|l||}\hline $u_N$ &The cover $\pi$& The reduction
of\\{}&{}&
 the holomorphic\\{}&{}& {differential}\\ \hline $u_3$
&$\rule[-3mm]{0mm}{8mm}\wp(u) = -{{z^3 - 27 g_3}\over {9( z^2 - 3
g_2)}} $& ${{d\wp}\over{\wp'}} = -{{3 z}\over2}{{dz}\over{w}}$ \\
     \cline{2-3} \hline $u_4$
     &$\rule[-3mm]{0mm}{8mm}\wp(u)=e_j+{(z-3e_i+9e_k)^2(z-z_1(i))
(z-z_2(i))\over
     4(z+6e_i)(-2z+15e_i)^2}$ {}& ${{d\wp}\over{\wp'}} =
-(2z+3e_k){{dz}\over{w}}$ \\ {}
&$\rule[-3mm]{0mm}{8mm}\wp(u)=e_k+{(z-3e_i+9e_j)^2
(z-z_3(i))(z-z_4(i))\over 4(z+6e_i)(-2z+15e_i)^2}$&
{}\\ \cline{2-3}\hline $u_5$ &$\rule[-3mm]{0mm}{8mm}\wp(u)=e_j +
{(z-z_1(j))(z-z_2(j))(z-z_3(j))(z+15e_j)^2 \over
(5z^2+6ze_j+261e_j^2-108g_2)^2 }$ {}& ${{d\wp}\over{\wp'}} =
(5z+3e_j){{dz}\over2{w}}$ \\ \hline\end{tabular} \end{table}

\begin{table}[htpb] \caption{The second cover}
\begin{tabular}{||l|l||}\hline $u_N$&The cover $\widetilde\pi$\\\hline
$u_3$ &$\rule[-3mm]{0mm}{8mm}\widetilde\wp(u) =
-{9\over16}(4z^3-9g_2z+9g_3)$ \\
  \cline{2-2}\hline $u_4$
&$\rule[-3mm]{0mm}{8mm}\widetilde\wp(u)-\widetilde e_j =
-{9(z-z_2)(z-z_3)(z+4e_i-e_k)^2\over 4(z+6e_i)} $ \\
 \cline{2-2}\hline $u_5$ &$\rule[-3mm]{0mm}{8mm}\widetilde\wp(u) =
-{9P_5(z)\over 4(z-3e_i-9e_j)(z+6e_i+9e_j)} $\\
 {}&$P_5(z)=z^5+3e_i^4z^4-42e_i^2z^3+150e_je_kz^3+30e_i^2e_kz^2 $\\
{}&$+9(69e_i^4+625e_j^2e_k^2-221e_ig_3/2)z$\\
     {}&$-27e_i(11e_i^4-81e_ig_k/2+475e_k^2e_j^2)$\\ \hline
\end{tabular} \end{table}

\begin{table}[htpb] \caption{The link between moduli of the tori $C_1$
and $\widetilde C_1$} \begin{tabular}{||l|l|l||}\hline
$u_3$&$\rule[-3mm]{0mm}{8mm}\widetilde g_2={3^7(g_2^3+9g_3^2)\over
2^6},\quad \widetilde g_3={3^{11}(g_3g_2^3-3g_3^3)\over 2^9} $\\
{}&$\rule[-3mm]{0mm}{8mm}{1\over \widetilde k}={1\over2}+
{(k^2-2)(2k^2-1)(k^2+1)\over4(k^4-k^2+1)^{3/2}}$\\ \hline $u_4$&$
\rule[-3mm]{0mm}{8mm}\widetilde
g_2=3^7{83\over4}g_2g_3e_i+3^6{89\over4}
g_2^2e_i^2+3^4g_2^2+14\cdot3^7g_3^2, \quad$\\ ${}$&$\widetilde g_3=
{3^7\cdot11\cdot17\over4}e_ig_2^4+{3^{10}\cdot
307\over8}e_ig_3^2g_2+{3^{10}\cdot457\over16}g_2^2g_3e_i^2
+{3^8\cdot61\over4}g_3g_2^3+{3^{11}\cdot5\cdot19\over16}g_3^3$\\
{}&$\rule[-3mm]{0mm}{8mm} \widetilde k=k'(1-4k^2) \quad {\rm or\;
equivalently}\; \widetilde k'=k(1-4(k')^2)$\\\hline $u_5$&$
\rule[-3mm]{0mm}{8mm}\widetilde g_2= 2^2 \cdot 3^7 \cdot 5 \cdot 23
e_i^6 - 3^7\cdot 11\cdot 17 e_i^4  g_2 - \frac{3^6 \cdot 5^3}{2^2}
e_i^2 g_2^2 +\frac {3^4}{ 2^4} 5^5 g_2^3 $\\
${}$&$\rule[-3mm]{0mm}{8mm}\widetilde g_3= - \frac {3^7}{2^6} e_i \left
(5^7g_2^4+ 2^8\cdot 3^5\cdot  191 e_i^8+ 2^5 \cdot
 3^3 \cdot  13 \cdot 457 e_i^4 g_2^2\right.$\\
${}$&$\rule[-3mm]{0mm}{8mm} \left.\qquad - 2^8\cdot 3^3 \cdot 23\cdot
79 e_i^6 g_2   -2^4 \cdot 3^2 \cdot 5^4 \cdot 11 e_i^2 g_2^3\right )
$\\\hline \end{tabular}\end{table}

\begin{table}[htpb] \caption{The wave function and Lam\'e polynomials}
\begin{tabular}{||l|l||}\hline $u_N$&The wave function $\Psi(x,u)$ and
Lam\'e polynomials $\Lambda(x)$ \\ \hline $u_3$
&$\rule[-3mm]{0mm}{8mm}\Psi(x)={\partial\over\partial x}({\rm exp}
(\lambda x)\Phi(x;u))$ \\
      &$ \Lambda_{ij}=\sqrt{(\wp(x)-e_i)(\wp(x)-e_j)}, (z=3e_k)\quad
    i\neq j\neq k=1,2,3
$\\{}&$\rule[-3mm]{0mm}{8mm}\Lambda_{\pm}=\wp(x)\pm {1\over
2}\sqrt{g_2\over3},(z=\pm\sqrt{3g_2})$\\ \cline{2-2} \hline $u_4$
&$\rule[-3mm]{0mm}{8mm}\Psi(x,u)={\partial\over\partial x}({\rm exp}
(\lambda x)\Phi(x;u)) +{3\lambda^2-3\wp(u)+z\over
6\sqrt{\wp(u)-e_i}}\Phi(x+\omega_i;u){\rm exp}(\lambda x)$ \\
&$\Lambda_{ik}^{\pm}, k\in \{1,2,3\}-\{i\}$\\ &$
\Lambda_{ik}=\sqrt{(\wp(x)-e_i)(\wp(x)-e_j)}+{1\over3}[(e_i-e_k)
\pm\sqrt{(e_i-e_k)(7e_i+e_k)}]\sqrt{\wp(x)-e_j\over \wp(x)-e_i}
$\\{}&$\rule[-3mm]{0mm}{8mm}\Lambda_{0}=\wp(x)-e_i$\\ \cline{2-2}
\hline $u_5$ &$\rule[-3mm]{0mm}{8mm}\Psi(x,u)={\partial\over\partial
x}({\rm exp} (\lambda x)\Phi(x;u))
+[a_{i,j}\Phi(x+\omega_i,u)+a_{j,i}\Phi(x+\omega_j,u)]{\rm exp}(\lambda
x)$ \\ &$a_{i,j}={-9\lambda\wp(u)+\lambda z+6\lambda e_i+3\wp'(u)\over
6\lambda\sqrt{\wp(u)-e_i}+6\sqrt{\wp(u)-e_j}\sqrt{\wp(u)-e_k}}$\\
&$\Lambda_i=\sqrt{(\wp(x)-e_i)(\wp(x)-e_k)}+(e_i-e_j)
{\sqrt{(\wp(x)-e_k)\over(\wp(x)-e_i)}}\,(z=6e_i-3e_j)$\\
&$\Lambda_i=\sqrt{(\wp(x)-e_j)(\wp(x)-e_k)}+
(e_j-e_i){\sqrt{(\wp(x)-e_k)\over(\wp(x)-e_j)}}\,(z=6e_j-3e_i)$\\
{}&$\Lambda_n=\sqrt{(\wp(x)-e_j)(\wp(x)-e_j)}+ \tilde a_{ij}
{\sqrt{(\wp(x)-e_i)\over(\wp(x)-e_j)}}+\tilde a_{ji}
{\sqrt{(\wp(x)-e_j)\over(\wp(x)-e_i)}}, $\\
{}&$\rule[-3mm]{0mm}{8mm}(z=z_k), i\neq j\neq k, \tilde a_{ij}
={15e_i^2+27e_j^2-6e_ie_j-z_n^2+2z_n(e_j-e_i)\over
24(e_j-e_i)}$\\\hline \end{tabular}\end{table}

\section{Theta Functional Formulae}

\subsection{Relations Between Theta Constants for $g=2$}
Here we give three groups of formulae which are consequence of the
Riemann theta formula  for theta constants when $g=2$.
These are the relations between the fourth powers of even theta
constants, the relations between the squares of even
theta constants and the Rosenhain derivative formulae.

\begin{eqnarray}
\theta^4\ma{0}{0}{0}{0}-\theta^4\ma{0}{0}{1}{1}&=&
\theta^4\ma{1}{0}{0}{1}+\theta^4\ma{0}{1}{0}{0}
=\theta^4\ma{1}{0}{0}{0}+\theta^4\ma{0}{1}{1}{0},\nonumber\\
\theta^4\ma{0}{0}{0}{0}-\theta^4\ma{1}{1}{0}{0}&=&
\theta^4\ma{0}{1}{1}{0}+\theta^4\ma{0}{0}{0}{1}
=\theta^4\ma{0}{0}{1}{0}+\theta^4\ma{1}{0}{0}{1},\nonumber\\
\theta^4\ma{0}{0}{0}{0}-\theta^4\ma{1}{1}{1}{1}&=&
\theta^4\ma{1}{0}{0}{0}+\theta^4\ma{0}{0}{1}{0}
=\theta^4\ma{0}{1}{0}{0}+\theta^4\ma{0}{0}{0}{1},\nonumber
\end{eqnarray}

\begin{eqnarray}
&&\theta^2\ma{0}{0}{0}{0}\theta^2\ma{1}{0}{0}{0}=
\theta^2\ma{0}{1}{0}{0}\theta^2\ma{1}{1}{0}{0}+
\theta^2\ma{0}{0}{0}{1}\theta^2\ma{1}{0}{0}{1},
\quad\left(\ma{1}{0}{0}{0}\right);\nonumber\\
&&\theta^2\ma{0}{0}{0}{0}\theta^2\ma{0}{1}{0}{0}=
\theta^2\ma{1}{0}{0}{0}\theta^2\ma{1}{1}{0}{0}+
\theta^2\ma{0}{1}{1}{0}\theta^2\ma{0}{0}{1}{0},
\quad\left(\ma{0}{1}{0}{0}\right);\nonumber\\
&&\theta^2\ma{0}{0}{0}{0}\theta^2\ma{1}{1}{0}{0}=
\theta^2\ma{1}{0}{0}{0}\theta^2\ma{0}{1}{0}{0}+
\theta^2\ma{0}{0}{1}{1}\theta^2\ma{1}{1}{1}{1},
\quad\left(\ma{1}{1}{0}{0}\right);\nonumber\\
&&\theta^2\ma{0}{0}{0}{0}\theta^2\ma{0}{0}{1}{0}=
\theta^2\ma{0}{1}{0}{0}\theta^2\ma{0}{1}{1}{0}+
\theta^2\ma{0}{0}{0}{1}\theta^2\ma{0}{0}{1}{1},
\quad\left(\ma{0}{0}{1}{0}\right);\nonumber\\
&&\theta^2\ma{1}{0}{0}{0}\theta^2\ma{0}{0}{1}{0}=
\theta^2\ma{0}{0}{1}{1}\theta^2\ma{1}{0}{0}{1}+
\theta^2\ma{0}{1}{1}{0}\theta^2\ma{1}{1}{0}{0},
\quad\left(\ma{1}{0}{1}{0}\right);\nonumber\\
&&\theta^2\ma{0}{1}{0}{0}\theta^2\ma{0}{0}{1}{0}=
\theta^2\ma{0}{0}{0}{0}\theta^2\ma{0}{1}{1}{0}+
\theta^2\ma{1}{1}{1}{1}\theta^2\ma{1}{0}{0}{1},
\quad\left(\ma{0}{1}{1}{0}\right);\nonumber\\
&&\theta^2\ma{0}{0}{1}{0}\theta^2\ma{1}{1}{0}{0}=
\theta^2\ma{1}{0}{0}{0}\theta^2\ma{0}{1}{1}{0}+
\theta^2\ma{1}{1}{1}{1}\theta^2\ma{0}{0}{0}{1},
\quad\left(\ma{1}{1}{1}{0}\right);\nonumber\\
&&\theta^2\ma{0}{0}{0}{0}\theta^2\ma{0}{0}{0}{1}=
\theta^2\ma{1}{0}{0}{0}\theta^2\ma{1}{0}{0}{1}+
\theta^2\ma{0}{0}{1}{1}\theta^2\ma{0}{0}{1}{0},
\quad\left(\ma{0}{0}{0}{1}\right);\nonumber\\
&&\theta^2\ma{1}{0}{0}{0}\theta^2\ma{0}{0}{0}{1}=
\theta^2\ma{0}{0}{0}{0}\theta^2\ma{1}{0}{1}{0}+
\theta^2\ma{1}{1}{1}{1}\theta^2\ma{0}{1}{1}{0},
\quad\left(\ma{1}{0}{0}{1}\right);\nonumber\\
&&\theta^2\ma{0}{1}{0}{0}\theta^2\ma{0}{0}{0}{1}=
\theta^2\ma{0}{0}{1}{1}\theta^2\ma{0}{1}{1}{0}+
\theta^2\ma{1}{0}{0}{1}\theta^2\ma{1}{1}{0}{0},
\quad\left(\ma{0}{1}{0}{1}\right);\nonumber\\
&&\theta^2\ma{1}{1}{0}{0}\theta^2\ma{0}{0}{0}{1}=
\theta^2\ma{1}{0}{0}{1}\theta^2\ma{0}{1}{0}{1}+
\theta^2\ma{0}{0}{1}{0}\theta^2\ma{1}{1}{1}{1},
\quad\left(\ma{1}{1}{0}{1}\right);\nonumber\\
&&\theta^2\ma{0}{0}{0}{0}\theta^2\ma{0}{0}{1}{1}=
\theta^2\ma{1}{1}{0}{0}\theta^2\ma{1}{1}{1}{1}+
\theta^2\ma{0}{0}{1}{0}\theta^2\ma{0}{0}{0}{1},
\quad\left(\ma{0}{0}{1}{1}\right);\nonumber\\
&&\theta^2\ma{0}{0}{1}{1}\theta^2\ma{1}{0}{0}{0}=
\theta^2\ma{0}{1}{0}{0}\theta^2\ma{1}{1}{1}{1}+
\theta^2\ma{1}{0}{0}{1}\theta^2\ma{0}{0}{1}{0},
\quad\left(\ma{1}{0}{1}{1}\right);\nonumber\\
&&\theta^2\ma{0}{0}{1}{1}\theta^2\ma{0}{1}{0}{0}=
\theta^2\ma{1}{0}{0}{0}\theta^2\ma{1}{1}{1}{1}+
\theta^2\ma{0}{1}{1}{0}\theta^2\ma{0}{0}{0}{1},
\quad\left(\ma{0}{1}{1}{1}\right);\nonumber\\
&&\theta^2\ma{1}{1}{0}{0}\theta^2\ma{0}{0}{1}{1}=
\theta^2\ma{0}{0}{0}{0}\theta^2\ma{1}{1}{1}{1}+
\theta^2\ma{1}{0}{0}{1}\theta^2\ma{0}{1}{1}{0},
\quad\left(\ma{1}{1}{1}{1}\right).\nonumber
\end{eqnarray}

We denote $D([\varepsilon],[\delta])=\theta_1[\varepsilon]\theta_2
[\delta]-\theta_2[\varepsilon]\theta_1[\delta]$. Then the following
Rosenhain formulae are valid

\begin{eqnarray}
&& D\left(
\left[\begin{array}{ll}\scriptstyle{0}&\!\!\!\!\scriptstyle{1}\cr
\scriptstyle{0}&\!\!\!\!\scriptstyle{1}\end{array}\right],
\left[\begin{array}{ll}\scriptstyle{1}&\!\!\!\!\scriptstyle{1}\cr
\scriptstyle{0}&\!\!\!\!\scriptstyle{1}\end{array}\right]
\right)=\pi^2
\theta\ma{0}{0}{1}{0}\theta\ma{0}{0}{1}{1}\theta\ma{1}{1}{1}{1}
\theta\ma{0}{1}{1}{0},\quad\left(\ma{1}{0}{0}{0}\right);
\nonumber\\
&&D\left(
\left[\begin{array}{ll}\scriptstyle{1}&\!\!\!\!\scriptstyle{1}\cr
\scriptstyle{1}&\!\!\!\!\scriptstyle{0}\end{array}\right],
\left[\begin{array}{ll}\scriptstyle{1}&\!\!\!\!\scriptstyle{0}\cr
\scriptstyle{1}&\!\!\!\!\scriptstyle{0}\end{array}\right]
\right)=\pi^2
\theta\ma{1}{1}{1}{1}\theta\ma{0}{0}{0}{1}\theta\ma{0}{0}{1}{1}
\theta\ma{1}{0}{0}{1},\quad\left(\ma{0}{1}{0}{0}\right);\nonumber\\
&&D\left(
\left[\begin{array}{ll}\scriptstyle{1}&\!\!\!\!\scriptstyle{0}\cr
\scriptstyle{1}&\!\!\!\!\scriptstyle{1}\end{array}\right],
\left[\begin{array}{ll}\scriptstyle{0}&\!\!\!\!\scriptstyle{1}\cr
\scriptstyle{1}&\!\!\!\!\scriptstyle{1}\end{array}\right]
\right)=\pi^2
\theta\ma{0}{1}{1}{0}\theta\ma{1}{0}{0}{1}\theta\ma{0}{0}{1}{0}
\theta\ma{0}{0}{0}{1},\quad\left(\ma{1}{1}{0}{0}\right);\nonumber\\
&&D\left(
\left[\begin{array}{ll}\scriptstyle{0}&\!\!\!\!\scriptstyle{1}\cr
\scriptstyle{1}&\!\!\!\!\scriptstyle{1}\end{array}\right],
\left[\begin{array}{ll}\scriptstyle{0}&\!\!\!\!\scriptstyle{1}\cr
\scriptstyle{0}&\!\!\!\!\scriptstyle{1}\end{array}\right]
\right)=\pi^2
\theta\ma{1}{0}{0}{0}\theta\ma{1}{0}{0}{1}\theta\ma{1}{1}{0}{0}
\theta\ma{1}{1}{1}{1},\quad\left(\ma{0}{0}{1}{0}\right);\nonumber\\
&&D\left(
\left[\begin{array}{ll}\scriptstyle{0}&\!\!\!\!\scriptstyle{1}\cr
\scriptstyle{1}&\!\!\!\!\scriptstyle{1}\end{array}\right],
\left[\begin{array}{ll}\scriptstyle{1}&\!\!\!\!\scriptstyle{1}\cr
\scriptstyle{0}&\!\!\!\!\scriptstyle{1}\end{array}\right]
\right)=\pi^2
\theta\ma{0}{0}{0}{0}\theta\ma{0}{0}{0}{1}\theta\ma{1}{1}{1}{1}
\theta\ma{0}{1}{0}{0},\quad\left(\ma{1}{0}{1}{0}\right);\nonumber\\
&&D\left(
\left[\begin{array}{ll}\scriptstyle{1}&\!\!\!\!\scriptstyle{0}\cr
\scriptstyle{1}&\!\!\!\!\scriptstyle{1}\end{array}\right],
\left[\begin{array}{ll}\scriptstyle{1}&\!\!\!\!\scriptstyle{1}\cr
\scriptstyle{0}&\!\!\!\!\scriptstyle{1}\end{array}\right]
\right)=\pi^2
\theta\ma{1}{1}{0}{0}\theta\ma{0}{0}{1}{1}\theta\ma{0}{0}{0}{1}
\theta\ma{1}{0}{0}{0},\quad\left(\ma{0}{1}{1}{0}\right);\nonumber\\
&&D\left(
\left[\begin{array}{ll}\scriptstyle{1}&\!\!\!\!\scriptstyle{0}\cr
\scriptstyle{1}&\!\!\!\!\scriptstyle{1}\end{array}\right],
\left[\begin{array}{ll}\scriptstyle{0}&\!\!\!\!\scriptstyle{1}\cr
\scriptstyle{0}&\!\!\!\!\scriptstyle{1}\end{array}\right]
\right)=\pi^2
\theta\ma{0}{1}{0}{0}\theta\ma{1}{0}{0}{1}\theta\ma{0}{0}{0}{0}
\theta\ma{0}{0}{1}{1},\quad\left(\ma{1}{1}{1}{0}\right);\nonumber\\
&&D\left(
\left[\begin{array}{ll}\scriptstyle{1}&\!\!\!\!\scriptstyle{0}\cr
\scriptstyle{1}&\!\!\!\!\scriptstyle{0}\end{array}\right],
\left[\begin{array}{ll}\scriptstyle{1}&\!\!\!\!\scriptstyle{0}\cr
\scriptstyle{1}&\!\!\!\!\scriptstyle{1}\end{array}\right]
\right)=\pi^2
\theta\ma{0}{1}{0}{0}\theta\ma{0}{1}{1}{0}\theta\ma{1}{1}{1}{1}
\theta\ma{1}{1}{0}{0},\quad\left(\ma{0}{0}{0}{1}\right);\nonumber\\
&&D\left(
\left[\begin{array}{ll}\scriptstyle{1}&\!\!\!\!\scriptstyle{1}\cr
\scriptstyle{1}&\!\!\!\!\scriptstyle{0}\end{array}\right],
\left[\begin{array}{ll}\scriptstyle{0}&\!\!\!\!\scriptstyle{1}\cr
\scriptstyle{1}&\!\!\!\!\scriptstyle{1}\end{array}\right]
\right)=\pi^2
\theta\ma{0}{0}{1}{1}\theta\ma{0}{0}{1}{0}\theta\ma{1}{1}{0}{0}
\theta\ma{0}{1}{0}{0},\quad\left(\ma{1}{0}{0}{1}\right);\nonumber\\
&&D\left(
\left[\begin{array}{ll}\scriptstyle{1}&\!\!\!\!\scriptstyle{1}\cr
\scriptstyle{1}&\!\!\!\!\scriptstyle{0}\end{array}\right],
\left[\begin{array}{ll}\scriptstyle{1}&\!\!\!\!\scriptstyle{0}\cr
\scriptstyle{1}&\!\!\!\!\scriptstyle{1}\end{array}\right]
\right)=\pi^2
\theta\ma{1}{1}{1}{1}\theta\ma{0}{0}{0}{0}\theta\ma{0}{0}{1}{0}
\theta\ma{1}{0}{0}{0},\quad\left(\ma{0}{1}{0}{1}\right);\nonumber\\
&&D\left(
\left[\begin{array}{ll}\scriptstyle{1}&\!\!\!\!\scriptstyle{0}\cr
\scriptstyle{1}&\!\!\!\!\scriptstyle{0}\end{array}\right],
\left[\begin{array}{ll}\scriptstyle{0}&\!\!\!\!\scriptstyle{1}\cr
\scriptstyle{1}&\!\!\!\!\scriptstyle{1}\end{array}\right]
\right)=\pi^2
\theta\ma{0}{1}{1}{0}\theta\ma{1}{0}{0}{0}\theta\ma{0}{0}{1}{1}
\theta\ma{0}{0}{0}{0},\quad\left(\ma{1}{1}{0}{1}\right);\nonumber\\
&&D\left(
\left[\begin{array}{ll}\scriptstyle{1}&\!\!\!\!\scriptstyle{1}\cr
\scriptstyle{1}&\!\!\!\!\scriptstyle{0}\end{array}\right],
\left[\begin{array}{ll}\scriptstyle{1}&\!\!\!\!\scriptstyle{1}\cr
\scriptstyle{0}&\!\!\!\!\scriptstyle{1}\end{array}\right]
\right)=\pi^2
\theta\ma{1}{0}{0}{1}\theta\ma{1}{0}{0}{0}\theta\ma{0}{1}{1}{0}
\theta\ma{0}{1}{0}{0},\quad\left(\ma{0}{0}{1}{1}\right);\nonumber\\
&&D\left(
\left[\begin{array}{ll}\scriptstyle{1}&\!\!\!\!\scriptstyle{1}\cr
\scriptstyle{1}&\!\!\!\!\scriptstyle{0}\end{array}\right],
\left[\begin{array}{ll}\scriptstyle{0}&\!\!\!\!\scriptstyle{1}\cr
\scriptstyle{0}&\!\!\!\!\scriptstyle{1}\end{array}\right]
\right)=\pi^2
\theta\ma{0}{0}{0}{1}\theta\ma{0}{0}{0}{0}\theta\ma{1}{1}{0}{0}
\theta\ma{0}{1}{1}{0},\quad\left(\ma{1}{0}{1}{1}\right);\nonumber\\
&&D\left(
\left[\begin{array}{ll}\scriptstyle{1}&\!\!\!\!\scriptstyle{0}\cr
\scriptstyle{1}&\!\!\!\!\scriptstyle{0}\end{array}\right],
\left[\begin{array}{ll}\scriptstyle{1}&\!\!\!\!\scriptstyle{1}\cr
\scriptstyle{0}&\!\!\!\!\scriptstyle{1}\end{array}\right]
\right)=\pi^2
\theta\ma{1}{1}{0}{0}\theta\ma{0}{0}{1}{0}\theta\ma{0}{0}{0}{0}
\theta\ma{1}{0}{0}{1},\quad\left(\ma{0}{1}{1}{1}\right);\nonumber\\
&&D\left(
\left[\begin{array}{ll}\scriptstyle{1}&\!\!\!\!\scriptstyle{0}\cr
\scriptstyle{1}&\!\!\!\!\scriptstyle{0}\end{array}\right],
\left[\begin{array}{ll}\scriptstyle{0}&\!\!\!\!\scriptstyle{1}\cr
\scriptstyle{0}&\!\!\!\!\scriptstyle{1}\end{array}\right]
\right)=\pi^2
\theta\ma{0}{1}{0}{0}\theta\ma{1}{0}{0}{0}\theta\ma{0}{0}{0}{1}
\theta\ma{0}{0}{1}{0},\quad\left(\ma{1}{1}{1}{1}\right).
\nonumber\end{eqnarray}

\subsection{ Addition Theorem for Second--Order \newline Theta
Functions at $g=2$} Here we give the expanded forms of (\ref{add}).  We
introduce the notation $\hat\theta[\varepsilon]({\bf
z})=\theta[\varepsilon]({\bf z}|2\tau)$.

\begin{eqnarray} &&\theta\ma{0}{0}{0}{0}\theta\ma{0}{0}{0}{0}({\bf z})=
\hat\theta^2\ma{0}{0}{0}{0}({\bf z})+\hat\theta^2\ma{1}{1}{0}{0}({\bf
z})+ \hat\theta^2\ma{1}{0}{0}{0}({\bf
z})+\hat\theta^2\ma{0}{1}{0}{0}({\bf z}),\nonumber\\
&&\theta\ma{0}{0}{1}{1}\theta\ma{0}{0}{1}{1}({\bf z})=
\hat\theta^2\ma{0}{0}{0}{0}({\bf z})+\hat\theta^2\ma{1}{1}{0}{0}({\bf
z})- \hat\theta^2\ma{1}{0}{0}{0}({\bf
z})-\hat\theta^2\ma{0}{1}{0}{0}({\bf z}),\nonumber\\
&&\theta\ma{0}{0}{1}{0}\theta\ma{0}{0}{1}{0}({\bf z})=
\hat\theta^2\ma{0}{0}{0}{0}({\bf z})-\hat\theta^2\ma{1}{1}{0}{0}({\bf
z})- \hat\theta^2\ma{1}{0}{0}{0}({\bf
z})+\hat\theta^2\ma{0}{1}{0}{0}({\bf z}),\nonumber\\
&&\theta\ma{0}{0}{0}{1}\theta\ma{0}{0}{0}{1}({\bf z})=
\hat\theta^2\ma{0}{0}{0}{0}({\bf z})-\hat\theta^2\ma{1}{1}{0}{0}({\bf
z})+ \hat\theta^2\ma{1}{0}{0}{0}({\bf
z})-\hat\theta^2\ma{0}{1}{0}{0}({\bf z}).\nonumber\end{eqnarray}

\begin{eqnarray} &&\theta\ma{1}{1}{0}{0}\theta\ma{1}{1}{0}{0}({\bf z})=
2\hat\theta\ma{1}{1}{0}{0}({\bf z})\hat\theta\ma{0}{0}{0}{0}({\bf z})+
2\hat\theta\ma{1}{0}{0}{0}({\bf z})\hat\theta\ma{0}{1}{0}{0}({\bf z}),
\nonumber\\ &&\theta\ma{1}{1}{1}{1}\theta\ma{1}{1}{1}{1}({\bf z})=
2\hat\theta\ma{1}{1}{0}{0}({\bf z})\hat\theta\ma{0}{0}{0}{0}({\bf z})-
2\hat\theta\ma{1}{0}{0}{0}({\bf z})\hat\theta\ma{0}{1}{0}{0}({\bf
z}),\nonumber\\ &&\theta\ma{0}{1}{0}{0}\theta\ma{0}{1}{0}{0}({\bf z})=
2\hat\theta\ma{0}{1}{0}{0}({\bf z})\hat\theta\ma{0}{0}{0}{0}({\bf z})+
2\hat\theta\ma{1}{0}{0}{0}({\bf z})\hat\theta\ma{1}{1}{0}{0}({\bf
z}),\nonumber\\ &&\theta\ma{0}{1}{1}{0}\theta\ma{0}{1}{1}{0}({\bf z})=
2\hat\theta\ma{0}{1}{0}{0}({\bf z})\hat\theta\ma{0}{0}{0}{0}({\bf z})-
2\hat\theta\ma{1}{0}{0}{0}({\bf z})\hat\theta\ma{1}{1}{0}{0}({\bf
z}),\nonumber\\ &&\theta\ma{1}{0}{0}{0}\theta\ma{1}{0}{0}{0}({\bf z})=
2\hat\theta\ma{1}{0}{0}{0}({\bf z})\hat\theta\ma{0}{0}{0}{0}({\bf z})+
2\hat\theta\ma{0}{1}{0}{0}({\bf z})\hat\theta\ma{1}{1}{0}{0}({\bf
z}),\nonumber\\ &&\theta\ma{1}{0}{0}{1}\theta\ma{1}{0}{0}{1}({\bf z})=
2\hat\theta\ma{1}{0}{0}{0}({\bf z})\hat\theta\ma{0}{0}{0}{0}({\bf z})-
2\hat\theta\ma{0}{1}{0}{0}({\bf z})\hat\theta\ma{1}{1}{0}{0}({\bf z})
\nonumber\end{eqnarray}

\begin{eqnarray} &&\theta\ma{1}{1}{0}{0}\theta\ma{1}{1}{0}{1}({\bf z})=
2\hat\theta\ma{1}{0}{0}{1}({\bf z})\hat\theta\ma{0}{1}{0}{1}({\bf z})+
2\hat\theta\ma{1}{1}{0}{1}({\bf z})\hat\theta\ma{0}{0}{0}{1}({\bf
z}),\nonumber\\ &&\theta\ma{1}{1}{1}{1}\theta\ma{1}{1}{1}{0}({\bf z})=
2\hat\theta\ma{1}{0}{0}{1}({\bf z})\hat\theta\ma{0}{1}{0}{1}({\bf z})-
2\hat\theta\ma{1}{1}{0}{1}({\bf z})\hat\theta\ma{0}{0}{0}{1}({\bf
z}),\nonumber\\ &&\theta\ma{0}{1}{0}{0}\theta\ma{0}{1}{0}{1}({\bf z})=
2\hat\theta\ma{0}{1}{0}{1}({\bf z})\hat\theta\ma{0}{0}{0}{1}({\bf z})+
2\hat\theta\ma{1}{0}{0}{1}({\bf z})\hat\theta\ma{1}{1}{0}{1}({\bf
z}),\nonumber\\ &&\theta\ma{0}{1}{1}{0}\theta\ma{0}{1}{1}{1}({\bf z})=
2\hat\theta\ma{0}{1}{0}{1}({\bf z})\hat\theta\ma{0}{0}{0}{1}({\bf z})-
2\hat\theta\ma{1}{0}{0}{1}({\bf z})\hat\theta\ma{1}{1}{0}{1}({\bf
z}),\nonumber\\ &&\theta\ma{1}{0}{0}{0}\theta\ma{1}{0}{1}{0}({\bf z})=
2\hat\theta\ma{0}{0}{1}{0}({\bf z})\hat\theta\ma{1}{0}{1}{0}({\bf z})+
2\hat\theta\ma{1}{1}{1}{0}({\bf z})\hat\theta\ma{0}{1}{1}{0}({\bf
z}),\nonumber\\ &&\theta\ma{1}{0}{0}{1}\theta\ma{1}{0}{1}{1}({\bf z})=
2\hat\theta\ma{0}{0}{1}{0}({\bf z})\hat\theta\ma{1}{0}{1}{0}({\bf z})-
2\hat\theta\ma{1}{1}{1}{0}({\bf z})\hat\theta\ma{0}{1}{1}{0}({\bf
z})\nonumber\end{eqnarray}

\begin{eqnarray} &&\theta\ma{1}{1}{0}{0}\theta_k\ma{1}{1}{0}{1}=
2\hat\theta\ma{0}{0}{0}{1}\hat\theta_k\ma{1}{1}{0}{1}+
2\hat\theta\ma{1}{0}{0}{1}\hat\theta_k\ma{0}{1}{0}{1},\nonumber\\
&&\theta\ma{1}{1}{1}{1}\theta_k\ma{1}{1}{1}{0}=
2\hat\theta\ma{1}{0}{0}{1}\hat\theta_k\ma{0}{1}{0}{1}-
2\hat\theta\ma{0}{0}{0}{1}\hat\theta_k\ma{1}{1}{0}{1},\nonumber\\
&&\theta\ma{0}{1}{0}{0}\theta_k\ma{0}{1}{0}{1}=
2\hat\theta\ma{0}{0}{0}{1}\hat\theta_k\ma{0}{1}{0}{1}+
2\hat\theta\ma{1}{0}{0}{1}\hat\theta_k\ma{1}{1}{0}{1},\nonumber\\
&&\theta\ma{0}{1}{1}{0}\theta_k\ma{0}{1}{1}{1}=
2\hat\theta\ma{0}{0}{0}{1}\hat\theta_k\ma{0}{1}{0}{1}-
2\hat\theta\ma{1}{0}{0}{1}\hat\theta_k\ma{1}{1}{0}{1},\nonumber\\
&&\theta\ma{1}{0}{0}{0}\theta_k\ma{1}{0}{1}{0}=
2\hat\theta\ma{0}{0}{1}{0}\hat\theta_k\ma{1}{0}{1}{0}+
2\hat\theta\ma{0}{1}{1}{0}\hat\theta_k\ma{1}{1}{1}{0},\nonumber\\
&&\theta\ma{1}{0}{0}{1}\theta_k\ma{1}{0}{1}{1}=
2\hat\theta\ma{0}{0}{1}{0}\hat\theta_k\ma{1}{0}{1}{0}-
2\hat\theta\ma{0}{1}{1}{0}\hat\theta_k\ma{1}{1}{1}{0}\nonumber
\end{eqnarray}

\subsection{Theta Constants of $2^p$-Sheeted Coverings over a
Torus}

In this subsection we denote  the  Jacobi theta constants by
$\vartheta_j=\vartheta_j\left(0|2^p\tau_{11}\right)$,
$\widetilde{\vartheta}_j=\vartheta_j\left(0|2^p\tau_{22}\right)$,
$j=2,3,4$.

{\bf p=1} Let $\tau=\left(\begin{array}{ll}\tau_{11}&{1\over2}\\
{1\over2}&\tau_{22}\end{array}\right)$
 Then \begin{eqnarray}&&\theta\ma{1}{0}{0}{0}=\theta\ma{1}{0}{0}{1}=
(2\vartheta_2\vartheta_3\widetilde{\vartheta}_3\widetilde{\vartheta}_4
)^{1/2},\, \theta\ma{0}{1}{1}{0}=\theta\ma{0}{1}{0}{0}=
(2\vartheta_3\vartheta_4\widetilde{\vartheta}_2\widetilde{\vartheta}_3
)^{1/2},\nonumber\\&&\hskip 28mm
\theta\ma{1}{1}{0}{0}=-i\theta\ma{1}{1}{1}{1}=
(2\vartheta_2\vartheta_4\widetilde{\vartheta}_2\widetilde{\vartheta}_4
)^{1/2},\nonumber\\ &&\theta\ma{0}{0}{0}{0}=
(\vartheta_3^2\widetilde{\vartheta}_3^2+\vartheta_2^2
\widetilde{\vartheta}_4^2+\vartheta_4^2\widetilde{\vartheta}_2^2)
^{1/2},\, \theta\ma{0}{0}{1}{1}=
(\vartheta_3^2\widetilde{\vartheta}_3^2-\vartheta_2^2
\widetilde{\vartheta}_4^2-\vartheta_4^2\widetilde{\vartheta}_2^2)
^{1/2},\nonumber\\ &&\theta\ma{0}{0}{1}{0}=
(\vartheta_3^2\widetilde{\vartheta}_3^2-\vartheta_2^2
\widetilde{\vartheta}_4^2+\vartheta_4^2\widetilde{\vartheta}_2^2)
^{1/2},\, \theta\ma{0}{0}{0}{1}=
(\vartheta_3^2\widetilde{\vartheta}_3^2+\vartheta_2^2
\widetilde{\vartheta}_4^2-\vartheta_4^2\widetilde{\vartheta}_2^2)
^{1/2},\nonumber\end{eqnarray}

\begin{eqnarray}&&\theta_1\ma{1}{1}{1}{0}=-{\pi}\theta\ma{1}{1}{0}{0}
\vartheta_3^2,\quad \theta_2\ma{1}{1}{1}{0}=-i\pi\theta\ma{1}{1}{1}{1}
\widetilde{\vartheta}_3^2,\nonumber\\
&&\theta_1\ma{1}{1}{0}{1}=-i\pi\theta\ma{1}{1}{0}{0}
\vartheta_3^2,\quad \theta_2\ma{1}{1}{0}{1}=-\pi\theta\ma{1}{1}{0}{0}
\widetilde{\vartheta}_3^2,\nonumber\\
&&\theta_1\ma{0}{1}{0}{1}=-i\pi\theta\ma{0}{1}{0}{0}
\vartheta_2^2,\quad \theta_2\ma{0}{1}{0}{1}=-\pi\theta\ma{0}{1}{0}{1}
\widetilde{\vartheta}_4^2,\nonumber\\
&&\theta_1\ma{0}{1}{1}{1}=i\pi\theta\ma{0}{1}{1}{0} \vartheta_2^2,\quad
\theta_2\ma{0}{1}{1}{1}=-\pi\theta\ma{0}{1}{1}{0}
\widetilde{\vartheta}_4^2,\nonumber\\
&&\theta_1\ma{1}{0}{1}{1}=-\pi\theta\ma{1}{0}{0}{1} \vartheta_4^2,\quad
\theta_2\ma{1}{0}{1}{1}=i\pi\theta\ma{1}{0}{0}{1}
\widetilde{\vartheta}_2^2,\nonumber\\
&&\theta_1\ma{1}{0}{1}{0}=-\pi\theta\ma{1}{0}{0}{0} \vartheta_4^2,\quad
\theta_2\ma{1}{0}{1}{0}=-i\pi\theta\ma{1}{0}{0}{0}
\widetilde{\vartheta}_2^2.\nonumber\end{eqnarray}

{\bf p=2} Let $\tau=\left(\begin{array}{ll}\tau_{11}&{1\over4}\\
{1\over4}&\tau_{22}\end{array}\right)$ and denote
$X=\vartheta_3\widetilde{\vartheta}_3$,
$Y=\vartheta_2\widetilde{\vartheta}_4$,
$Z=\vartheta_4\widetilde{\vartheta}_2$,
$A=-X^2+Y^2+Z^2$, $B=X^2-Y^2+Z^2$, $C=X^2+Y^2-Z^2$, $D=A+B+C$.
Then the following formulae hold:

\begin{eqnarray}
&&\theta\ma{0}{0}{0}{0}=X+Y+Z,\quad\theta\ma{0}{0}{0}{1}=X+Y-Z,\nonumber\\
&&\theta\ma{0}{0}{1}{0}=X-Y+Z,\quad\theta\ma{0}{0}{1}{1}=X-Y-Z,\nonumber\\
&&\theta^2\ma{1}{0}{0}{0}=2^{3/2}(XY)^{1/2}(D^{1/2}+2^{1/2}Z),\,
\theta^2\ma{1}{0}{0}{1}=2^{3/2}(XY)^{1/2}(D^{1/2}-2^{1/2}Z),\nonumber\\
&&\theta^2\ma{0}{1}{0}{0}=2^{3/2}(XZ)^{1/2}(D^{1/2}+2^{1/2}Y),\,
\theta^2\ma{0}{1}{1}{0}=2^{3/2}(XZ)^{1/2}(D^{1/2}-2^{1/2}Y),\nonumber\\
&&\theta^2\ma{1}{1}{0}{0}=2^{3/2}(YZ)^{1/2}(D^{1/2}+2^{1/2}X),
\theta^2\ma{1}{1}{1}{1}=2^{3/2}(YZ)^{1/2}(D^{1/2}-2^{1/2}X).
\nonumber\end{eqnarray}

\begin{eqnarray} &&\theta_1\ma{1}{0}{1}{0}={ -\pi}(2XY)^{1/4}
(\vartheta_4^2B^{1/2}+2^{1/2}\vartheta_3^2Z)
(D^{1/2}+2^{1/2}Z)^{-1/2},\nonumber\\
&&\theta_2\ma{1}{0}{1}{0}=-{i\pi}(2XY)^{1/4}
(\widetilde{\vartheta}_2^2B^{1/2}+2^{1/2}\widetilde{\vartheta}_3^2Z)
(D^{1/2}+2^{1/2}Z)^{-1/2},\nonumber\\
&&\theta_1\ma{0}{1}{0}{1}={{-i}\pi}(2XZ)^{1/4}
({\vartheta}_2^2C^{1/2}+2^{1/2}{\vartheta}_3^2Y)
(D^{1/2}+2^{1/2}Y)^{-1/2},\nonumber\\
&&\theta_2\ma{0}{1}{0}{1}=-{\pi}(2XZ)^{1/4}
(\widetilde{\vartheta}_4^2C^{1/2}+2^{1/2}\widetilde{\vartheta}_3^2Y)
(D^{1/2}+2^{1/2}Y)^{-1/2},\nonumber\\ &&\theta_1\ma{1}{1}{0}{1}={
-i\pi}(2ZY)^{1/4} ({\vartheta}_3^2C^{1/2}+2^{1/2}{\vartheta}_2^2X)
 (D^{1/2}+2^{1/2}X)^{-1/2},\nonumber\\
&&\theta_2\ma{1}{1}{0}{1}=-{\pi}(2ZY)^{1/4}
(\widetilde{\vartheta}_3^2C^{1/2}+2^{1/2}\widetilde{\vartheta}_4^2X)
(D^{1/2}+2^{1/2}X)^{-1/2},\nonumber\\
&&\theta_1\ma{1}{1}{1}{0}=-{\pi}(2ZY)^{1/4}
({\vartheta}_3^2B^{1/2}+2^{1/2}{\vartheta}_4^2X)
 (D^{1/2}+2^{1/2}X)^{-1/2},\nonumber\\
&&\theta_2\ma{1}{1}{1}{0}={-\pi}(2ZY)^{1/4}
(\widetilde{\vartheta}_3^2B^{1/2}+2^{1/2}\widetilde{\vartheta}_4^2X)
(D^{1/2}+2^{1/2}X)^{-1/2},\nonumber\\
&&\theta_1\ma{1}{0}{1}{1}={{-i\pi}}(2XY)^{1/4}
(\vartheta_4^2A^{1/2}-2^{1/2}i\vartheta_2^2Z)
 (D^{1/2}+2^{1/2}Z)^{-1/2},\nonumber\\ &&\theta_2\ma{1}{0}{1}{1}=-{
i\pi}(2XY)^{1/4}
(\widetilde{\vartheta}_2^2A^{1/2}-2^{1/2}i\widetilde{\vartheta}_3^2Z)
(D^{1/2}+2^{1/2}Z)^{-1/2},\nonumber\\
&&\theta_1\ma{0}{1}{1}{1}={\pi}(2XZ)^{1/4}
({\vartheta}_2^2A^{1/2}-2^{1/2}i{\vartheta}_3^2Y)
(D^{1/2}+2^{1/2}Y)^{-1/2},\nonumber\\
&&\theta_2\ma{0}{1}{1}{1}=-i\pi(2XZ)^{1/4}
(\widetilde{\vartheta}_4^2A^{1/2}-2^{1/2}i\widetilde{\vartheta}_3^2Y)
(D^{1/2}+2^{1/2}Y)^{-1/2}.\nonumber\end{eqnarray}

\end{document}